\documentclass[11pt,a4paper,final]{article}
\usepackage{amsmath}%
\numberwithin{equation}{section}
\usepackage{amstext}%
\usepackage{amssymb}%
\usepackage{showkeys}%
\usepackage{epsfig}%
\usepackage{graphicx}%
\usepackage{xcolor}
\usepackage{multicol}
\usepackage{amsthm}
\usepackage{booktabs}
\usepackage{apacite}
\usepackage{multirow}
\usepackage[top=1in, bottom=1.25in, left=0.8in, right=0.8in]{geometry}
\DeclareMathOperator*{\plim}{plim}

\theoremstyle{prop}
\theoremstyle{proof}

\providecommand{\keywords}[1]{
\textbf{Keywords:~~~} Stochastic DDM, ML estimator, Bayesian estimator, Kalman filtering, Regime--switching.
}

\begin{document}
\title{Parameter Estimation Methods of Required Rate of Return on Stock}
\author{Battulga Gankhuu\footnote{Department of Applied Mathematics, National University of Mongolia; E-mail: battulga.g@seas.num.edu.mn; Phone Number: 976--99246036}}
\date{}

\maketitle 

\begin{abstract}
In this study, we introduce new estimation methods for the required rate of return of the stochastic dividend discount model (DDM) and the private company valuation model, which will appear below. To estimate the required rate of return, we use the maximum likelihood method, the Bayesian method, and the Kalman filtering. We apply the model to a set of firms from the S\&P 500 index using historical dividend and price data over a 32--year period. Overall, suggested methods can be used to estimate the required rate of return. Suggested methods not only used to estimate required rate of return on stock, but also used to estimate required rate of return of debtholders.
\end{abstract}


\section{Introduction}

Dividend discount models (DDMs), first introduced by \citeA{Williams38}, are a popular tool for stock valuation. If we assume that a firm will not default in the future, then the basic idea of all DDMs is that the market price of a stock of a firm equals the sum of the stock's future dividends discounted at risk--adjusted rates (required rate of return on stock) 
\begin{equation}\label{eq01}
P_t=\sum_{m=1}^\infty \frac{d_{t+m}}{\prod_{j=1}^m(1+k_{t+j}^\circ)},
\end{equation}
where $P_t$ is the stock price, $d_t$ is the dividend payment, and $k_t^\circ$ is the required rate of return on a stock, respectively at time $t$ of a company. By their very nature, DDM approaches are best applicable to companies paying regular cash dividends. For a DDM with default risk, we refer to \citeA{Battulga22a}. From equation \eqref{eq01}, it is obvious fact that in addition to dividend forecast models, the required rate of return is the main input parameter for DDMs. In addition to its usage in stock valuation,  it is an ingredient of the weighted average cost of capital (WACC), and WACC is used to value businesses and projects, see \citeA{Brealey20}. The most common model to estimate the required rate of return is the capital asset pricing model (CAPM). Using the CAPM is common in practice, but it is a one--factor model ($\beta$ only) for which criticism applies, see, e.g., \citeA{Nagorniak85}. Thus, multi--factor models (e.g., \citeA{Fama93}) are therefore often preferred instead. Another multi--factor model, which is used to estimate the required rate of return is \citeauthor{Ross76}'s \citeyear{Ross76} arbitrage pricing theory (APT). However, for the model, since every analyst can develop his APT model, there is no universally accepted APT model specification among practitioners.  On the other hand, as the outcome of DDMs depends crucially on dividend forecasts, most research in the last few decades has been around the proper estimations of dividend development. An interesting review of some existing deterministic and stochastic DDMs, which model future dividends can be found in \citeA{dAmico20}. 

Sudden and dramatic changes in the financial market and economy are caused by events such as wars, market panics, or significant changes in government policies. To model those events, some authors used regime--switching models. The regime--switching model was introduced by seminal works of \citeA{Hamilton89,Hamilton90} (see also a book of \citeA{Hamilton94}) and the model is a hidden Markov model with dependencies, see \citeA{Zucchini16}. The regime--switching model assumes that a discrete unobservable Markov process switches among a finite set of regimes randomly and that each regime is defined by a particular parameter set. The model is a good fit for some financial data and becomes popular in financial modeling including equity options, bond prices, and others.

The Kalman filtering, which was introduced by \citeA{Kalman60} is an algorithm that provides estimates of some unobserved (state) vector, which is driven by a stochastic process given the measurements, which is observed over time. The Kalman filtering has been demonstrating its usefulness in various applications. It has been used extensively in economics, system theory, the physical sciences, and engineering. In econometrics, the state--space model is usually defined by (i) the observed vector is described in terms of the state vector in linear form (measurement equation), and (ii) the state vector is governed by VAR(1) process (transition equation). To estimate the parameters of the state--space model and to make inferences about the state--space model (smoothing and forecasting), the Kalman filtering can be used, see \citeA{Hamilton94} and \citeA{Lutkepohl05}.

As mentioned the above most popular practical method, which is used to estimate the required rate of return is the CAPM. By the CAPM, the required rate of return is modeled by the risk--free rate, beta, and market return. However, the CAPM is sensitive to its inputs. Recently, \citeA{Battulga22a} introduced a stochastic DDM that models the dividends by a compound non--homogeneous Poisson--process and obtained ML estimators and confidence bands of the model's parameters, including the required rate of return. In this paper, instead of the traditional CAPM and its descendant versions, we introduce new estimation methods, which cover the ordinary ML methods, the Bayesian method, ML methods with regime--switching, and the Kalman filtering to estimate the required rate of return and we extend the result, which is related to the required rate of return of \citeA{Battulga22a}.

The rest of the paper is organized as follows: In Section 2, we introduce estimation methods of the required rate of return of the stochastic DDM using the ordinary ML method and ML method with regime--switching. Section 3 is devoted to parameter estimations of a private company valuation model, where we consider the ordinary ML method, the Bayesian method, ML method with regime--switching, and the Kalman filtering to estimate the parameters. In Section 4, for selected public companies, we provide numerical results based on our methods. Finally, Section 5 concludes the study.

\section{Parameters Estimation of Public Company}

In this paper, we assume that a company will not default in the future. Due to equation \eqref{eq01}, for successive prices of the company, the following relation holds 
\begin{equation}\label{eq02}
P_t=(1+k_t^\circ)P_{t-1}-d_t,~~~t=1,2,\dots,
\end{equation}
where $k_t^\circ$ is the required rate of return, $P_t$ is the stock price, and $d_t$ is the dividend, respectively, at time $t$ of the company. In order to obtain parameter estimators, we assume that at time $t$, the stock price differs from its theoretical value by a random amount, say, $u_t$, we have observations up to and including time $T$ of prices and dividends. Under this assumption equation \eqref{eq02} becomes 
\begin{equation}\label{eq03}
P_t=(1+k_t^\circ)P_{t-1}-d_t+u_t,~~~t=1,\dots,T,
\end{equation}
Throughout the paper, we assume that $\{u_t\}$ is a sequence of independent identically normally distributed random variables with means of zero and variances of $\sigma^2$, that is, $u_t\sim \mathcal{N}(0,\sigma^2)$. From equation \eqref{eq02}, one can conclude that for the fixed value of dividend at time $t$, the stock price at time $t$ is explained by the required rate of return at time $t$. Therefore, one should model the required rate of return. As the required rate of return at time $t$ may depend on macroeconomic variables and firm--specific variables, such as GDP, inflation, key financial ratios of the firm, and so on, we model the required rate of return by the following linear equation
\begin{equation}\label{eq04}
k_t^\circ=k_1+k_2c_{2t}+\dots+k_nc_{nt}=c_t'k,~~~t=1,\dots,T,
\end{equation}
where $k:=(k_1,\dots,k_n)'$ is an $(n\times 1)$ parameter vector of the required rate of return, $c_{it}$ indicates $i$--th covariate at time $t$, and $c_t:=(c_{1t},c_{2t},\dots,c_{nt})'$ is an $(n\times 1)$ covariate vector at time $t$, which is used to explain the required rate of return with $c_{1t}=1$. For a special case corresponding to the required rate of return is constant, namely, $k_t^\circ=k$, we refer to \citeA{Battulga22a}. In this section, we assume that the required rate of return takes a positive value, that is, $k_t^\circ>0$ for all $t=1,\dots,T$. If we combine equations \eqref{eq03} and \eqref{eq04}, then our model becomes
\begin{equation}\label{eq05}
P_t=(1+c_t'k)P_{t-1}-d_t+u_t,~~~t=1,\dots,T.
\end{equation}
To keep notations simple, let us introduce the following vectors and matrix: $p:=(p_1,\dots,p_T)'$ is a $(T\times 1)$ price vector, $d:=(d_1,\dots,d_T)'$ is a $(T\times 1)$ dividend vector, $u:=(u_1,\dots,u_T)'$ is a $(T\times 1)$ random error vector, $p_{-1}:=(p_0,\dots,p_{T-1})'$ is a $(T\times 1)$ lagged price vector, and $X':=[x_1:\dots:x_T]$ is an $(n\times T)$ matrix, which consists of the covariates and prices, and whose $t$--th column is $x_t:=c_tP_{t-1}$. Then, the model \eqref{eq05} has a simple representation, namely,
\begin{equation}\label{eq06}
p=p_{-1}+Xk-d+u.
\end{equation}
The representation implies that the log--likelihood function of the model \eqref{eq06} is given by  
\begin{equation}\label{eq07}
\mathcal{L}(\theta)=-\frac{T}{2}\ln(\sigma^2)-\frac{1}{2\sigma^2}\Big(p+d-p_{-1}-Xk\Big)'\Big(p+d-p_{-1}-Xk\Big),
\end{equation}
where $\theta:=(k',\sigma^2)'$ is an $((n+1)\times 1)$ parameter vector. Taking partial derivatives from the log--likelihood with respect to the parameters and setting these partial derivatives to zero gives the maximum likelihood estimators as
\begin{equation}\label{eq08}
\hat{k}:=(X'X)^{-1}X'(p+d-p_{-1})~~~\text{and} ~~~\hat{\sigma}^2:=\frac{1}{T}e'e,
\end{equation}
where $e:=p+d-p_{-1}-X\hat{k}$ is a $(T\times 1)$ unrestricted residual vector, and $\hat{k}$ and $\hat{\sigma}^2$ are unrestricted maximum likelihood estimators of the parameters $k$ and $\sigma^2$, respectively. It should be noted that if we take $d=0$ in equation \eqref{eq08}, then we obtain the estimators that correspond to a non--dividend paying company. If we substitute equation \eqref{eq06} into equation \eqref{eq08} of estimator $\hat{k}$, one obtains that
\begin{equation}\label{eq09}
\hat{k}=k+(X'X)^{-1}X'u.
\end{equation}
Substituting equation \eqref{eq09} into the unrestricted residual vector $e$, we obtain that the unrestricted residual vector is represented by $e=Mu$, where $M_X:=I_T-X(X'X)^{-1}X'$ is a symmetric idempotent matrix. Therefore, an unrestricted residual sum of squares is given by
\begin{equation}\label{eq10}
e'e=u'M_Xu.
\end{equation}

Because the required rate of return takes a positive value, that is, $k_t^\circ=c_t'k>0$, the price process $P_t$ is explosive. Therefore, to test the significance of the parameters of the required rate of return, one can not use the usual $t$ statistic directly. Due to this reason, we aim to obtain asymptotic distributions of the parameters $k_1,\dots,k_n$. For the rest of the section, we follow the ideas of \citeA{Hasza77} and \citeA{Johnston97}. Equation \eqref{eq05}, which begins at time zero can be written as
\begin{equation}\label{}
P_t=\prod_{m=1}^t(1+c_m'k)P_0+\sum_{m=1}^t\bigg(\prod_{j=m+1}^t(1+c_j'k)\bigg)\big(u_m-d_m\big).
\end{equation}
Let for $t=1,\dots,T$, $\tau_t:=\prod_{m=1}^t(1+c_m'k)$ be an accumulated value over a time interval $[0,t]$ of one U.S. dollar at the required rate of returns $c_1'k,\dots,c_t'k$ and $\rho_t=1/\tau_t$ be a discount factor, which is a reciprocal of the accumulated value over a time interval $[0,t]$. Then, the random price at time $t$ of the stock is represented by
\begin{equation}\label{eq11}
P_t=\tau_t\xi_t,
\end{equation}
where $\xi_t:=P_0+\sum_{m=1}^t\rho_m(u_m-d_m).$ If we assume that $\sum_{m=1}^\infty\rho_m|d_m|<\infty$, then by the Beppo--Levi theorem (see \citeA{Capinski04}) one deduce that
\begin{equation}\label{eq12}
\xi_t\overset{\text{a.s.}}{\longrightarrow}\xi:=\sum_{m=1}^\infty\rho_m u_m,
\end{equation}
where a.s. is an abbreviation of the almost sure convergence. Therefore, for a residual term of the random variable $\xi$, we have
\begin{equation}\label{eq13}
\xi-\xi_t=\sum_{m=t+1}^\infty\rho_m(u_m-d_m)=O_p(\rho_t).
\end{equation}
Let $\Psi':=[\psi_1:\dots:\psi_T]$ be an $(n\times T)$ nonrandom matrix, whose $t$--th column is an $(n\times 1)$ vector $\psi_t:=(\psi_{1t},\dots,\psi_{nt})'$ with $\psi_{it}:=c_{it}(\rho_T\tau_{t-1})$. This nonrandom matrix will play a crucial role in this section, see below. Let us denote $(i,j)$--th element of a generic matrix $A$ and $i$--th element of a generic vector $a$ by $(A)_{ij}$ and $(a)_i$, respectively. Then, since $(X'X)_{ij}=\sum_{t=1}^Tc_{it}c_{jt}P_{t-1}^2$ and $(X'u)_i=\sum_{t=1}^Tc_{it}P_{t-1}u_t$, according to equations \eqref{eq11} and \eqref{eq13}, one obtain that
\begin{eqnarray}\label{eq14}
\rho_T^2(X'X)_{ij}&=&\rho_T^2\sum_{t=1}^Tc_{it}c_{jt}\tau_{t-1}^2\big(\xi^2+2\xi(\xi_{t-1}-\xi)+(\xi_{t-1}-\xi)^2\big)\nonumber\\
&=&z^2\sum_{t=1}^Tc_{it}c_{jt}(\rho_T\tau_{t-1})^2+O_p(\rho_T)=z^2(\Psi'\Psi)_{ij}+O_p(\rho_T)
\end{eqnarray}
and 
\begin{eqnarray}\label{eq15}
\rho_T(X'u)_i&=&(z\rho_T)\sum_{t=1}^Tc_{it}(\rho_T\tau_{t-1})u_t+\rho_T\sum_{t=1}^Tc_{it}\tau_{t-1}(\xi_{t-1}-\xi)u_t\nonumber\\
&=&z\sum_{t=1}^Tc_{it}(\rho_T\tau_{t-1})u_t+O_p(T\rho_T)=z(\Psi'u)_i+O_p(T\rho_T).
\end{eqnarray}
Since due to equation \eqref{eq12} and the structure of the matrix $\Psi'$, we have $\xi=\sum_{t=1}^{[T/2]}\rho_t u_t+O_p(\rho_{[T/2]})$ and $(\Psi'u)_i=\sum_{t=[T/2]+1}^Tc_{it}(\rho_T\tau_{t-1})u_t+O_p(\rho_{[T/2]})$ for $i=1,\dots,q$, the random variable $\xi$ and random vector $\Psi' u$ are asymptotically independent, where $[x]$ is an integer part of a real number $x\in \mathbb{R}$. Note that the matrix $\Psi$ is nonrandom. Then, according to equations \eqref{eq09}, \eqref{eq14}, and \eqref{eq15}, we get that
\begin{equation}\label{eq16}
\tau_T(\hat{k}-k)=(\rho_T^2X'X)^{-1}(\rho_TX'u)=\frac{1}{\xi}(\Psi'\Psi)^{-1}\Psi'u+O_p(T\rho_T).
\end{equation}
Therefore, $\hat{k}$ is a consistent estimator. It follows from equations \eqref{eq10}, \eqref{eq15}, and \eqref{eq16} that the unconstrained residual sum of squares can be represented by
\begin{equation}\label{}
e'e=u'u-\tau_T(\hat{k}-k)'(\rho_TX'u)=u'u-u'\Psi(\Psi'\Psi)^{-1}\Psi' u+O_p(T\rho_T)=u'M_\Psi u+O_p(T\rho_T),
\end{equation}
where $M_\Psi:=I_T-\Psi(\Psi'\Psi)^{-1}\Psi'$ is a symmetric idempotent matrix with a rank $\text{rank}(M_\Psi)=\text{tr}(M_\Psi)=T-n$. Thus, as $u\sim \mathcal{N}(0,\sigma^2 I_T)$, a ratio of the unconstrained residual sum of squares to the parameter $\sigma^2$, $e'e/\sigma^2$, follows approximately a chi-squared distribution with $(T-n)$ degrees of freedom, that is, 
\begin{equation}\label{eq17}
\frac{e'e}{\sigma^2} \approx \chi^2(T-n)
\end{equation}
where $\approx$ means approximately holds. Therefore, $\hat{\sigma}^2$ is a consistent estimator, that is, p$\lim\hat{\sigma}^2=\sigma^2$, where $\plim$ is the probability limit. By equations \eqref{eq14} and \eqref{eq16} we get that
$$\frac{1}{\hat{\sigma}}(X'X)^{\frac{1}{2}}(\hat{k}-k)=\frac{\text{sign}(\xi)}{\hat{\sigma}}(\Psi'\Psi)^{-\frac{1}{2}}\Psi u+O_p(T\rho_T)$$
As a result, because the random variable $\xi$ and the random vector $\Psi' u$ are asymptotically independent, and $\hat{\sigma}^2$ is the consistent estimator, we have
\begin{equation}\label{eq87}
\frac{1}{\hat{\sigma}_u}(X'X)^{\frac{1}{2}}(\hat{k}-k)\overset{d}{\longrightarrow}\mathcal{N}(0,I_n),
\end{equation}
where $d$ means convergence in distribution. Therefore, it follows from equation \eqref{eq87} that the parameter estimator $\hat{k}$ follows approximately multivariate normal distribution with mean $k$ and covariance matrix $\sigma^2(X'X)^{-1}$, that is,
\begin{equation}\label{eq18}
\hat{k}\approx\mathcal{N}\big(k,\sigma^2(X'X)^{-1}\big).
\end{equation}
Now, let us consider the following general linear hypothesis
\begin{equation}\label{eq19}
H_0: \mathsf{R}k=r,
\end{equation} 
where $\mathsf{R}$ is a $(q\times n)$ known matrix with rank $q$, with $q<n$, and $r$ is a ($q\times 1$) known vector. It should be noted that the general linear hypothesis includes the following hypotheses: ($i$) $H_0: k_i=0$ is may be used to test $i$--the covariate has no influence to explain the stock price and the test is referred to as a significance test, ($ii$) $H_0: k_i=k_{i*}$ is may be used to test the parameter $k_i$ equals to some specific value $k_{i*}$, $(iii)$ $H_0: k_2=\dots=k_n=0$ is may be used to test all the covariates jointly do not influence to explain the stock price, $(iv)$ $H_0: k_1+k_2c_{2*}+\dots+k_nc_{n*}=k_*^\circ$ is may be used to test a linear combination of the parameters $k_1,\dots,k_n$ is equals to some specific value ($k_*^\circ$) of the required rate of return, where $c_{2*},\dots,c_{n*}$ are some given values, and so on, see \citeA{Johnston97}. 

It can be shown that if the general linear hypothesis \eqref{eq19} is true, then it holds
\begin{equation}\label{eq20}
\big(\mathsf{R}\hat{k}-r\big)'\big[\sigma^2\mathsf{R}(X'X)^{-1}\mathsf{R}'\big]^{-1}\big(\mathsf{R}\hat{k}-r\big)\approx \chi^2(q).
\end{equation}
Because the matrices $M_\Psi$ and $(\Psi'\Psi)^{-1}\Psi$ are orthogonal, the estimator $\hat{k}$ and the residual sum of squares $e'e$ are asymptotically independent. Consequently, if we combine equations \eqref{eq17} and \eqref{eq20}, then we obtain that
\begin{equation}\label{eq21}
F=\big(\mathsf{R}\hat{k}-r\big)'\big[s^2\mathsf{R}(X'X)^{-1}\mathsf{R}'\big]^{-1}\big(\mathsf{R}\hat{k}-r\big)/q\approx F(q,T-n),
\end{equation} 
where $s^2=\frac{e'e}{T-n}$ is an estimator, which is asymptotically equivalent to the estimator $\hat{\sigma}^2$ and $F(m,n)$ is the Fisher's $F$ distribution with $(m,n)$ degrees of freedom. Note that for the $F$ statistic, $qF$ statistic converges in distribution to a chi--squared random variable with $q$ degrees of freedom, that is, $qF\overset{d}{\longrightarrow}\chi^2(q)$ as $T\to\infty$. If the value of the $F$ statistic is greater than $F_{1-\alpha}(q,T-n)$, then we reject the hypothesis \eqref{eq19} at significance level $\alpha$, where $F_{1-\alpha}(q,T-n)$ is a $(1-\alpha)$ quantile of the $F$ distribution with $(q,T-n)$ degrees of freedom.

Let us consider a special case, which is practically useful for the general linear hypothesis \eqref{eq19}, namely, for $i=1,\dots,n$, $H_0: k_i=k_{i*}$. Hypothesis $H_0: k_i=k_{i*}$ is tested by the following $t$ statistic
\begin{equation}\label{}
t=\frac{\hat{k}_i-k_{i*}}{s\sqrt{\big((X'X)^{-1}\big)_{ii}}}\approx t(T-n).
\end{equation}
If the absolute value of the computed $t$ statistic exceeds the $t_{1-\alpha/2}(T-n)$, that is, $|t|>t_{1-\alpha/2}(T-n)$, then we reject the hypothesis $H_0: k_i=k_{i*}$, where $t_{1-\alpha/2}(T-n)$ is a $(1-\alpha/2)$ quantile of the student $t$ distribution with $(T-n)$ degrees of freedom. For $i=1,\dots,n$, $(1-\alpha)100\%$ confidence interval of the parameter $k_i$ is obtained by
\begin{equation}\label{}
\hat{k}_i-t_{1-\alpha/2}(T-n) s\sqrt{\big((X'X)^{-1}\big)_{ii}}\leq k_i\leq \hat{k}_i+t_{1-\alpha/2}(T-n) s\sqrt{\big((X'X)^{-1}\big)_{ii}}.
\end{equation}
In particular, if the required rate of return is constant, that is, $k_t^\circ=k$, then the confidence interval becomes
\begin{equation}\label{eq88}
\hat{k}-\frac{t_{1-\alpha/2}(T-1) s}{\sqrt{p_{-1}^Tp_{-1}}}\leq k\leq \hat{k}+\frac{t_{1-\alpha/2}(T-1) s}{\sqrt{p_{-1}^Tp_{-1}}}.
\end{equation}
For a confidence interval of the constant required rate of return based on the Likelihood Ratio (LR) statistic, we refer to \citeA{Battulga22a}. In practice, instead of statistic \eqref{eq21}, to test the general linear hypothesis it is often preferred to use a simple statistic, which is only based on the unrestricted residual sum of squares $e'e$ and a restricted residual sum of squares $e_*'e_*$, defined below. To obtain the statistic we need to consider the following constrained optimization problem 
\begin{equation}\label{eq22}
\begin{cases}
\mathcal{L}(\theta)\longrightarrow \max\\
\text{s.t.}~\mathsf{R}k=r
\end{cases},
\end{equation}
where $\mathcal{L}(\theta)$ is the log--likelihood function given by equation \eqref{eq07}. It can be shown that restricted estimators corresponding to the constrained optimization problem are given by
\begin{equation}\label{eq23}
\hat{k}_*=\hat{k}-(X'X)^{-1}\mathsf{R}'[\mathsf{R}(X'X)^{-1}\mathsf{R}']^{-1}(\mathsf{R}\hat{k}-r)~~~\text{and}~~~\hat{\sigma}_*^2=\frac{1}{T}e_*'e_*,
\end{equation}
where $e_*:=p+d-p_{-1}-X\hat{k}_*=e+X(\hat{k}-\hat{k}_*)$ is a $(T\times 1)$ restricted residual vector, and $\hat{k}_*$ and $\hat{\sigma}_*^2$ are restricted maximum likelihood estimators of the parameters $k$ and $\sigma^2$, respectively. As $X'e=0$, the restricted sum of squares is 
\begin{equation}\label{eq24}
e_*'e_*=e'e+(\hat{k}-\hat{k}_*)'X'X(\hat{k}-\hat{k}_*).
\end{equation}
If we substitute equation \eqref{eq23} into equation \eqref{eq24}, one gets that
\begin{equation}\label{}
e_*'e_*-e'e=\big(\mathsf{R}\hat{k}-r\big)'\big[\mathsf{R}(X'X)^{-1}\mathsf{R}'\big]^{-1}\big(\mathsf{R}\hat{k}-r\big)+O_p(T\rho_T).
\end{equation}
Consequently, by substituting the above equation into equation \eqref{eq21}, one can obtain the following statistic, which is used to test general linear hypothesis \eqref{eq19} and depends only on the unrestricted sum of squares $e'e$ and the restricted sum of squares $e_*'e_*$:
\begin{equation}\label{}
F=\frac{(e_*'e_*-e'e)/q}{e'e/(T-n)}\approx F(q,T-n).
\end{equation}
Since the unrestricted parameter estimator $\hat{k}$ is consistent, it follows from equations \eqref{eq23} and \eqref{eq24} that the restricted estimators $\hat{k}_*$ and $\hat{\sigma}_*^2$ are consistent.

Now, to test general linear hypothesis \eqref{eq19}, we consider the LR statistic, which is based on the maximum likelihood method. If hypothesis $H_0$ is true, then it follows from equation \eqref{eq09} that $\mathsf{R}\hat{k}-r=\mathsf{R}(X'X)^{-1}X'u$. Therefore, according to equation \eqref{eq23}, we have
\begin{equation}\label{eq90}
\hat{k}_*=\hat{k}-(X'X)^{-1}\mathsf{R}'[\mathsf{R}(X'X)^{-1}\mathsf{R}']^{-1}\mathsf{R}(X'X)^{-1}X'u.
\end{equation}
As a result, it follows from equation \eqref{eq24} that
\begin{equation}\label{}
e_*'e_*-e'e=u'\bar{M}_Xu,
\end{equation}
where $\bar{M}_X:=X(X'X)^{-1}\mathsf{R}'[\mathsf{R}(X'X)^{-1}\mathsf{R}']^{-1}\mathsf{R}(X'X)^{-1}X'$ is a symmetric idempotent matrix with a rank $\text{rank}(\bar{M}_X)=\text{tr}(\bar{M}_X)=q$. It can be shown that the LR statistic corresponding to the general linear hypothesis is given by
\begin{equation}\label{eq25}
\text{LR}=2\big[\mathcal{L}(\hat{\theta})-\mathcal{L}(\hat{\theta}_*)\big]=T\ln\bigg(1+\frac{e_*'e_*-e'e}{e'e}\bigg),
\end{equation}
where $\hat{\theta}:=(\hat{k}',\hat{\sigma}^2)'$ is a vector of unrestricted estimators, $\hat{\theta}_*:=(\hat{k}_*',\hat{\sigma}_{u*}^2)'$ is a vector of restricted estimators, $\mathcal{L}(\hat{\theta})$ is an unrestricted maximum value of the log--likelihood function, and $\mathcal{L}(\hat{\theta}_*)$ is a restricted maximum value of the log--likelihood function subject to the constraint $\mathsf{R}\beta=r$. Using equations \eqref{eq14} and \eqref{eq15} one obtain that a difference between the restricted and the unrestricted residual sum of squares $e_*'e_*-e'e$ is represented by
\begin{equation}\label{eq26}
e_*'e_*-e'e=u'\bar{M}_Xu=u'\bar{M}_\Psi u+O_p(T\rho_T),
\end{equation}
where $\bar{M}_\Psi:=\Psi(\Psi'\Psi)^{-1}\mathsf{R}'[\mathsf{R}(\Psi'\Psi)^{-1}\mathsf{R}']^{-1}\mathsf{R}(\Psi'\Psi)^{-1}\Psi'$ is a nonrandom, symmetric, and idempotent matrix whose rank is $\text{rank}(\bar{M})=\text{tr}(\bar{M})=q$. Therefore, a ratio of the right--hand side's first term of equation \eqref{eq26} to $\sigma^2$ follows the chi--squared distribution with $q$ degrees of freedom, that is, $u'\bar{M}_\Psi u/\sigma^2\sim \chi^2(q)$. Thus, from equation \eqref{eq17}, we have 
\begin{equation}\label{}
\plim_{T\to\infty}\bigg\{\frac{e_*'e_*-e'e}{e'e}\bigg\}=0.
\end{equation}
Consequently, according to equations \eqref{eq25} and \eqref{eq26}, for the LR statistic, the following approximation holds
\begin{equation}\label{eq89}
\text{LR}\approx\frac{T(e_*'e_*-e'e)}{e'e}=\frac{u'\bar{M}_\Psi u+O_p(T\rho_T)}{\hat{\sigma}^2},~~~T\to\infty.
\end{equation}
Because $\hat{\sigma}^2$ is the consistent estimator and $u'\bar{M}_\Psi u/\sigma^2$ follows the chi--squared distribution with $q$ degrees of freedom, the LR statistic converges in distribution to a random variable whose distribution is the chi--square with $q$ degrees of freedom, that is,
\begin{equation}\label{}
\text{LR}=T\ln\bigg(1+\frac{e_*'e_*-e'e}{e'e}\bigg)\overset{d}{\longrightarrow}\chi^2(q),
\end{equation}
where $e'e$ is the unrestricted residual sum of squares and $e_*'e_*$ is the restricted residual sum of squares. According to \citeA{Johnston97}, it can be shown that the Wald (W) and the Lagrangian Multiplier (LM) statistics are represented by the following equations
\begin{equation}\label{}
\text{W}=\frac{T(e_*'e_*-e'e)}{e'e} ~~~\text{and}~~~\text{LM}=\frac{T(e_*'e_*-e'e)}{e_*'e_*}
\end{equation}
respectively. Therefore, it follows from equation \eqref{eq89} and the fact that the restricted estimator $\hat{\sigma}_*^2$ is consistent that asymptotic distributions of the statistics are obtained by 
\begin{equation}\label{}
\text{W}=\frac{T(e_*'e_*-e'e)}{e'e}\overset{d}{\longrightarrow}\chi^2(q) ~~~\text{and}~~~\text{LM}=\frac{T(e_*'e_*-e'e)}{e_*'e_*}\overset{d}{\longrightarrow}\chi^2(q),
\end{equation}
respectively, see also \citeA{Lutkepohl05}. If the values of the LR, W, and LM statistics are greater than $\chi_{1-\alpha}^2(q)$, then we reject the hypothesis \eqref{eq19} at significance level $\alpha$, where $\chi_{1-\alpha}^2(q)$ is a $(1-\alpha)$ quantile of the chi--squared distribution with $q$ degrees of freedom. 

\subsection{Regime--Switching Estimation}

This section is devoted to regime--switching estimators of parameters of the required rate of return and is based on the book of \citeA{Hamilton94}. Let us consider DDM in \eqref{eq03}/\eqref{eq05} with $N$ regimes, which is given by the following equation 
\begin{equation}\label{eq38}
P_t=\big(1+k_t^\circ(s_t)\big)P_{t-1}-d_t+u_t=\big(1+c_t'k(s_t)\big)P_{t-1}-d_t+u_t,
\end{equation}
where $P_t$ is the stock price and $d_t$ is the dividend, respectively, at time $t$ of the company, $c_t$ is the $(n\times 1)$ vector of covariates at time $t$, $u_t$ is the random error at time $t$, $s_t$ is an unobserved regime at time $t$, which is governed by a Markov chain with $N$ states, and $k(s_t)$ is an $(n\times 1)$ parameter vector of the required rate of return corresponding to the regime $s_t$. The coefficient vector for this model is $k(1)$ when the process is in regime 1, $k(2)$ when the process is in regime 2, and so on. In this section, we assume that the regime--switching process $s_t$ is governed by first--order homogeneous Markov chain. This means a conditional probability that the regime at time $t$, $s_t$ equals some particular value conditional on the past regimes, $s_{t-1},s_{t-2},\dots,s_1$ depends only through the most recent regime at time $t-1$, $s_{t-1}$, and does not depend on time, that is, 
\begin{equation}\label{}
p_{ij}:=\mathbb{P}(s_t=j|s_{t-1}=i)=\mathbb{P}(s_t=j|s_{t-1}=i,s_{t-2}=r_{t-2},\dots,s_1=r_1),~~~i,j=1,\dots,N.
\end{equation}
If we collect all the conditional probabilities $p_{ij}$ into a matrix $\mathsf{P}$, then we obtain a transition probability matrix of the regime--switching process $s_t$
\begin{equation}\label{eq39}
\mathsf{P}=\begin{bmatrix}
p_{11} & p_{12} & \dots & p_{1N}\\
p_{21} & p_{22} & \dots & p_{2N}\\
\vdots & \vdots & \ddots & \vdots\\
p_{N1} & p_{N2} & \dots & p_{NN}\\
\end{bmatrix}.
\end{equation}
Observe that sums of all rows of the transition probability matrix $\mathsf{P}$ equals 1, that is, for all $i=1,\dots,N$, $p_{i1}+\dots+p_{iN}=1.$ For $t=0,\dots,T$, let us denote available information at time $t$ by $\mathcal{F}_t$, which consists of the prices, dividends, and covariates: $\mathcal{F}_t:=(P_0,P_1,\dots,P_t,d_1,\dots,d_t,c_1',\dots,c_t')'.$ Then, it is clear that the log--likelihood function of our model is given by the following equation
\begin{equation}\label{}
\mathcal{L}(\theta)=\sum_{t=1}^Tf(P_t|d_t,c_t,\mathcal{F}_{t-1};\theta)
\end{equation}
where $\theta:=\big(k(1)',\dots,k(N)',\sigma^2,\rho',\text{vec}(\mathsf{P})'\big)'$ is a vector, which consists of all population parameters of the model and $f(P_t|d_t,c_t,\mathcal{F}_{t-1};\theta)$ is a conditional density function of the price at time $t$, $P_t$. Here the $\text{vec}$ is an operator that transforms $\mathsf{P}$ into an $(N^2\times 1)$ vector by stacking the columns and $\rho:=(\mathbb{P}(s_1|\mathcal{F}_0),\dots,\mathbb{P}(s_N|\mathcal{F}_0))'$ is an $(N\times 1)$ initial probability vector. The log--likelihood is used to obtain the maximum likelihood estimator of the parameter vector $\theta$. Note that the log--likelihood function depends on all observations, which are collected in $\mathcal{F}_T$, but does not depend on regime--switching process $s_t$, whose values are unobserved. If we assume that the regime--switching process in regime $j$ at time $t$, then because $u_t$ follows a normal distribution with mean zero and variance $\sigma^2$, the conditional density function of the price at time $t$ of the company $P_t$ is given by the following equation
\begin{equation}\label{eq40}
\eta_{tj}:=f(P_t|s_t=j,d_t,c_t,\mathcal{F}_{t-1};\alpha)=\frac{1}{\sqrt{2\pi}\sigma}\exp\bigg\{-\frac{\big(P_t+d_t-\big(1+c_t'k(j)\big)P_{t-1}\big)^2}{2\sigma^2}\bigg\}
\end{equation}
for $t=1,\dots,T$ and $j=1,\dots,N$, where $\alpha:=(k(1)',\dots,k(N)',\sigma^2)'$ is a parameter vector, which differs from the vector of all parameters $\theta$ by the initial probability vector $\rho$ and transition probability matrix $\mathsf{P}$. For all $t=1,\dots,T$, we collect the conditional density functions of the price at time $t$ into an $(n\times 1)$ vector $\eta_t$, that is, $\eta_t:=(\eta_{t1},\dots,\eta_{tN})'$. Let us denote a probabilistic inference about the value of the regime--switching process $s_t$ equals to $j$, based on the information $\mathcal{F}_t$ and the parameter vector $\theta$ by $\mathbb{P}(s_t=j|\mathcal{F}_t,\theta)$. Collect these conditional probabilities $\mathbb{P}(s_t=j|\mathcal{F}_t,\theta)$ for $j=1,\dots,N$ into an $(N\times 1)$ vector $z_{t|t}$, that is, $z_{t|t}:=\big(\mathbb{P}(s_t=1|\mathcal{F}_t;\theta),\dots,\mathbb{P}(s_t=N|\mathcal{F}_t;\theta)\big)'$. Also, we need a probabilistic forecast about the value of the regime--switching process at time $t+1$ equals $j$ conditional on data up to and including time $t$. Collect these forecasts into an $(N\times 1)$ vector $z_{t+1|t}$, that is, $z_{t+1|t}:=\big(\mathbb{P}(s_{t+1}=1|\mathcal{F}_t;\theta),\dots,\mathbb{P}(s_{t+1}=N|\mathcal{F}_t;\theta)\big)'$.  

The probabilistic inference and forecast for each time $t=1,\dots,T$ can be found by iterating on the following pair of equations: 
\begin{equation}\label{eq41}
z_{t|t}=\frac{(z_{t|t-1}\odot\eta_t)}{i_N'(z_{t|t-1}\odot\eta_t)}~~~\text{and}~~~z_{t+1|t}=\mathsf{P}'z_{t|t},~~~t=1,\dots,T,
\end{equation}
where $\eta_t$ is the $(N\times 1)$ vector, whose $j$-th element is given by equation \eqref{eq40}, $\mathsf{P}$ is the $(N\times N)$ transition probability matrix, which is given by equation \eqref{eq39}, $i_N$ is an $(N\times 1)$ vector, whose elements are equal 1, and the $\odot$ is the Hadamard (element--wise) product. Given a starting value $\rho=z_{1|0}$ and an assumed value for the population parameter vector $\theta$, one can iterate on \eqref{eq41} for $t=1,\dots,T$ to calculate the values of $z_{t|t}$ and $z_{t+1|t}$. To obtain MLE of the population parameters, in addition to the inferences and forecasts we need a smoothed inference about the regime--switching process was in at time $t$ based on full information $\mathcal{F}_T$. Collect these smoothed inferences into an $(N\times 1)$ vector $z_{t|T}$, that is, $z_{t|T}:=\big(\mathbb{P}(s_t=1|\mathcal{F}_T;\theta),\dots,\mathbb{P}(s_t=N|\mathcal{F}_T;\theta)\big)'$. The smoothed inferences can be obtained by using the \citeauthor{Kim94}'s \citeyear{Kim94} smoothing algorithm:
\begin{equation}\label{eq42}
z_{t|T}=z_{t|t}\odot\big\{\mathsf{P}'(z_{t+1|T}\oslash z_{t+1|t})\big\},~~~t=T-1,\dots,1,
\end{equation}
where $\oslash$ is an element--wise division of two vectors. The smoothed probabilities $z_{t|T}$ are found by iterating on \eqref{eq42} backward for $t=T-1,\dots,1$. This iteration is started with $z_{T|T}$, which is obtained from \eqref{eq41} for $t=T$.

If the initial probability $\rho$ does not depend on the other parameters, then according to \citeA{Hamilton90}, maximum likelihood estimators of $(i,j)$-th element of the transition probability matrix $\mathsf{P}$, the parameter vector $\alpha$ that governs the conditional density functions \eqref{eq40}, and the initial probability $\rho$ are obtained from the following systems of equations
\begin{eqnarray}
\hat{p}_{ij}&=&\frac{\sum_{t=2}^T\mathbb{P}\big(s_{t-1}=i,s_t=j|\mathcal{F}_T;\hat{\theta}\big)}{\sum_{t=2}^T(z_{t-1|T})_i},\label{eq43}\\
0&=&\sum_{t=1}^T\bigg(\frac{\partial \ln(\eta_t)}{\partial\alpha'}\bigg)'z_{t|T},\label{eq44}\\
\hat{\rho}&=&z_{1|T},\label{eq45}
\end{eqnarray}
where ${\partial \ln(\eta_t)}/{\partial\alpha'}$ is an $(N \times (n+1))$ matrix of derivatives of the 
logs of the conditional densities and due to the Kim's smoothing algorithm, the numerator of equation \eqref{eq43} can be calculated by
\begin{equation}\label{}
\mathbb{P}\big(s_{t-1}=i,s_t=j|\mathcal{F}_T;\theta\big)=p_{ij}(z_{t|T})_j(z_{t-1|t-1})_i/{(z_{t|t-1})_j}.
\end{equation}
To simplify notations for MLE that corresponds to the parameter vector $\alpha$, let us define the following matrix and vectors: for each regime $j=1,\dots,N$, $\bar{X}_j':=\big[\bar{x}_{1,j}:\dots:\bar{x}_{T,j}\big]$ is an $(n\times T)$ matrix whose $t$--th column is given by an $(n\times 1)$ vector $\bar{x}_{t,j}:=x_t\sqrt{(z_{t|T})_j}$ and which is obtained by adjusting the matrix $X'$ by the regime $j$, and $\bar{p}_j:=(\bar{p}_{1,j},\dots,\bar{p}_{T,j})'$, $\bar{d}_j:=(\bar{d}_{1,j},\dots,\bar{d}_{T,j})'$, and $\bar{p}_{-1,j}:=(\bar{p}_{1,j}^*,\dots,\bar{p}_{T,j}^*)'$ are $(T\times 1)$ vectors whose $t$--th elements are given by $\bar{p}_{t,j}:=p_t\sqrt{(z_{t|T})_j}$, $\bar{d}_{t,j}:=d_t\sqrt{(z_{t|T})_j}$, and $\bar{p}_{t,j}^*:=p_{t-1}\sqrt{(z_{t|T})_j}$ and which are obtained by adjusting the vectors $p$, $d$, and $p_{-1}$ by the regime $j$, respectively. It follows from equation \eqref{eq44} that for the parameters $k(1),\dots,k(N)$ and $\sigma^2$, maximum likelihood estimators, which are analogies of equation \eqref{eq08}  are obtained by
\begin{equation}\label{eq46}
\hat{k}(j)=\big(\bar{X}_{j}'\bar{X}_j\big)^{-1}\bar{X}_j'\big(\bar{p}_j+\bar{d}_j-\bar{p}_{-1,j}\big)~~~\text{for}~j=1,\dots,N ~~~\text{and} ~~~\hat{\sigma}^2=\frac{1}{T}\sum_{j=1}^N\bar{e}_j'\bar{e}_j,
\end{equation}
where $\bar{e}_j:=\bar{p}_j+\bar{d}_j-\bar{p}_{-1,j}-\bar{X}_j\hat{k}(j)$ is a $(T\times 1)$ residual vector, which corresponds to $j$--th regime. The maximum likelihood estimator of the parameter vector $\theta$ is obtained by the zig--zag iteration method using the equations \eqref{eq41}--\eqref{eq43}, \eqref{eq45}, and \eqref{eq46}.

\section{Parameters Estimation of Private Company}

In this section, we will consider parameter estimation methods for a private company. Let $B_t$ be a book value of equity, $R_t$ be a return of equity (ROE), $b_t$ be a book value growth rate, and $\alpha_t$ be a payout ratio, respectively, at time $t$ of the private company. Since the book value of equity at time $t-1$ grows at rate $b_t$, its value at time $t$ becomes 
\begin{equation}\label{eq27}
B_t=(1+b_t)B_{t-1}.
\end{equation}
On the other side, as dividend payment at time $t$ is a product of the dividend payout ratio at time $t$ and earning at time $t$, $R_tB_{t-1}$, we have 
\begin{equation}\label{}
d_t=\Delta_tB_{t-1},
\end{equation}
where $\Delta_t:=\alpha_tR_t=d_t/B_{t-1}$ is a dividend--to--book ratio at time $t$. If we assume that a price--to--book ratio is constant, say, $m=P_t/B_t$, for all $t=1,\dots,T$, then according to DDM equation \eqref{eq02}, price (value) at time $t$ of the company is expressed by the following equation
\begin{equation}\label{eq28}
mB_t=(1+k_t^\circ)mB_{t-1}-\alpha_tR_tB_{t-1}=\big((1+k_t^\circ)m-\Delta_t\big)B_{t-1},
\end{equation}
where $k_t^\circ$ is the required rate of return at time $t$ given by equation \eqref{eq04}. If we multiply equation \eqref{eq27} by the price--to--book ratio and equate it to the left--hand side of the above equation \eqref{eq28}, then we get that
\begin{equation}\label{}
(1+b_t)mB_{t-1}=\big((1+k_t^\circ)m-\Delta_t\big)B_{t-1}.
\end{equation}
Therefore, a relation between the dividend--to--book ratio, book value growth rate, required rate of return, and the price--to--book ratio is given by
\begin{equation}\label{ad01}
mb_t=m k_t^\circ+\Delta_t.
\end{equation}
We refer to the model and its versions with the regime--switching and with a state (latent or unobserved) variable given in equations \eqref{ad08} and \eqref{eq55}, respectively, as the private company valuation model. For the log private company valuation model, we refer to \citeA{Battulga22d}, where he considers the private company valuation model in the log framework, and obtain closed--form pricing and hedging formulas for European call and put options. It should be noted that the private company valuation model given in \eqref{} is equivalent to the franchise factor model, see \citeA{Leibowitz90}, but the private company valuation models with the regime--switching and the state variable differ from the franchise factor model. According to equation \eqref{ad01}, the required rate of return at time $t$ is represented by
\begin{equation}\label{eq29}
k_t^\circ=\frac{1}{m}\Delta_t+b_t.
\end{equation}
From the above equation, one can see that for a dividend--paying private company, if $m$ increases, then the required rate of return $k_t^\circ$ decreases and it converges to the book value growth rate $b_t$. Thus, as the price--to--book and dividend--to--book ratios are positive, the book value growth rate is a floor of the required rate of return. To estimate the parameters of the required rate of return and the price--to--book ratio, we must add a random amount, say, $u_t$, to equation \eqref{eq29}. Recall that $k_t^\circ=c_t'k$. Then, equation \eqref{eq29} becomes 
\begin{equation}\label{eq30}
b_t=c_t'k-\delta \Delta_t+u_t, ~~~t=1,\dots,T,
\end{equation}
where $\delta:=1/m$ is a book--to--price ratio. It should be noted that if the company does not pay dividends, then we must remove the term $\delta \Delta_t$ from equation \eqref{eq30}. In this case, equation \eqref{eq30} becomes
\begin{equation}\label{ad02}
b_t=c_t'k+u_t, ~~~t=1,\dots,T.
\end{equation}
Let us define the following vectors and matrix: $\Delta:=(\Delta_1,\dots,\Delta_T)'$ is a $(T\times 1)$ dividend--to--book ratio vector, $b:=(b_1,\dots,b_T)'$ is a $(T\times 1)$ book value growth rate vector, $u:=(u_1,\dots,u_T)'$ is a $(T\times 1)$ random error vector, and $C':=[c_1:\dots:c_T]$ is an $(n\times T)$ covariate matrix. Then, equations \eqref{eq30} and \eqref{ad02} can be written by
\begin{equation}\label{ad03}
b=Ck-\delta \Delta+u, ~~~u\sim\mathcal{N}(0,\sigma^2I_T)
\end{equation}
and
\begin{equation}\label{ad04}
b=Ck+u, ~~~u\sim\mathcal{N}(0,\sigma^2I_T),
\end{equation}
respectively. Further, we define: for the dividend--paying company, $\beta:=(k',\delta)'$ is an $((n+1)\times 1$) parameter vector, and $X:=[C:-\Delta]$ is a ($T\times (n+1)$) matrix, composed of all independent variables, and for the non--dividend paying company, $\beta:=k$ is an $(n\times 1)$ parameter vector, and $X:=C$ is a $(T\times n)$ matrix, composed of all covariates. Then, equations \eqref{ad03} and \eqref{ad04} have an even simple representation
\begin{equation}\label{eq31}
b=X\beta+u, ~~~u\sim\mathcal{N}(0,\sigma^2I_T).
\end{equation}
Then, the log--likelihood function of the private company valuation model is given by the following equation
\begin{equation}\label{eq32}
\mathcal{L}(\theta)=-\frac{T}{2}\ln(2\pi)-\frac{T}{2}\ln(\sigma^2)-\frac{1}{2\sigma^2}\Big(b-X\beta\Big)'\Big(b-X\beta\Big),
\end{equation}
where for the dividend--paying company, $\theta:=(k',\delta,\sigma^2)$ is an $((n+2)\times 1)$ parameter vector, which consists of all parameters of the model, and for the non--dividend paying company, $\theta:=(k',\sigma^2)$ is an $((n+1)\times 1)$ parameter vector, which also consists of all parameters of the model. From log--likelihood function \eqref{eq32}, one can obtain that maximum likelihood estimators of the model's parameters are given by
\begin{equation}\label{eq33}
\hat{\beta}=(X'X)^{-1}X'b~~~\text{and}~~~\hat{\sigma}^2=\frac{1}{T}e'e,
\end{equation}
where $e:=b-X\hat{\beta}$ is a $(T\times 1)$ unrestricted residual vector. Thus, a theoretical value at time $t$ of the dividend--paying company may be obtained by that its book value at time $t$ of equity multiplied by its parameter estimation of price--to--book ratio, namely, $\hat{m}=1/\hat{\delta}$. It is the well--known fact that the maximum likelihood estimator $\hat{\beta}$ equals the least square estimator of the parameter $\beta$ and the maximum likelihood estimator $\hat{\sigma}^2$ differs the least square estimator of the parameter $\sigma^2$ by a multiplier: for the dividend--paying company, $\frac{T}{T-n-1}$ and the non--dividend paying company, $\frac{T}{T-n}$. Note that for the dividend--paying company, the parameter estimator vector $\hat{\beta}$ can be decomposed by
\begin{equation}\label{}
\hat{\delta}=-\frac{\Delta'M_Cg}{\Delta'M_C\Delta}~~~\text{and}~~~\hat{k}=(C'C)^{-1}C'[\hat{\delta}\Delta+b],
\end{equation}
where $M_C:=I_T-C(C'C)^{-1}C'$ is a $(T\times T)$ symmetric idempotent matrix. It is easy to show that $\hat{\beta}$ is an unbiased and consistent estimator of parameter $\beta$, and $\hat{\sigma}^2$ is an asymptotically unbiased and consistent estimator of the $\sigma^2$. 

As mentioned the above one often needs to test the general linear hypothesis. Let us consider the following general linear hypothesis: for the dividend--paying company, 
\begin{equation}\label{eq36}
H_0: \mathsf{R}\beta=\mathsf{R}_1k+\mathsf{R}_2\delta=r,
\end{equation}
where $\mathsf{R}_1$ is a $(q\times n)$ $(q\leq n)$ known matrix, corresponding to the parameters of the required rate of return, $k$, $\mathsf{R}_2$ is a $(q\times 1)$ known vector, corresponding to the book--to--price ratio parameter, $\delta$, $r$ is a $q\times 1$ known vector, and $\mathsf{R}=[\mathsf{R}_1:\mathsf{R}_2]$ is a stacked matrix, and for the non--dividend paying company,
\begin{equation}\label{}
H_0: \mathsf{R}\beta=r,
\end{equation}
where $\mathsf{R}$ is a $(q\times n)$ $(q\leq n)$ known matrix, corresponding to the parameters of the required rate of return, $k$, and $r$ is a $q\times 1$ known vector. To test the hypothesis, we need to consider the following constrained optimization problem, which covers both dividend--paying and non--paying companies
\begin{equation}\label{}
\begin{cases}
\mathcal{L}(\theta)\longrightarrow \max\\
\text{s.t.}~\mathsf{R}\beta=r, 
\end{cases}
\end{equation}
where $\mathcal{L}(\theta)$ is the log--likelihood function given in \eqref{eq32} of the private company valuation model. Then, one can show that restricted maximum likelihood estimators, which is a solution of the above constrained optimization problem are obtained (see equation \eqref{eq23}) by  
\begin{equation}\label{}
\hat{\beta}_*=\hat{\beta}-(X'X)^{-1}\mathsf{R}'[\mathsf{R}(X'X)^{-1}\mathsf{R}']^{-1}(\mathsf{R}\hat{\beta}-r)~~~\text{and}~~~\hat{\sigma}_*^2=\frac{1}{T}e_*'e_*,
\end{equation}
where $e_*:=g-X\hat{\beta}_*$ is a $(T\times 1)$ restricted residual vector. Now we list statistics, used to test the hypothesis and construct confidence intervals for the parameters
\begin{itemize}
\item[1.] $F$ statistic:
\begin{equation}\label{}
F=\frac{(e_*'e_*-e'e)/q}{e'e/(n-k)}\sim F(q,n-k),
\end{equation}
\item[2.] LR statistic:
\begin{equation}\label{}
\text{LR}=T\ln\bigg(1+\frac{e_*'e_*-e'e}{e'e}\bigg)\approx \chi^2(q),
\end{equation}
\item[3.] Wald statistic:
\begin{equation}\label{}
\text{W}=\frac{T(e_*'e_*-e'e)}{e'e}\approx \chi^2(q),
\end{equation}
\item[4.] and Lagrangian multiplier (LM) statistic:
\begin{equation}\label{}
\text{LM}=\frac{T(e_*'e_*-e'e)}{e_*'e_*}\approx \chi^2(q),
\end{equation}
\end{itemize}
where $e'e$ is an unrestricted residual sum of squares and $e_*'e_*$ is a restricted residual sum of squares of the private company valuation model, for detailed analysis, see \citeA{Johnston97}. For the dividend--paying company, if the value of the $F$ statistic is greater than $F_{1-\alpha}(q,T-n-1)$, then we reject hypothesis \eqref{eq36} at significance level $\alpha$, where $F_{1-\alpha}(q,T-n-1)$ is the $(1-\alpha)$ quantile of the Fisher's $F$ distribution with $(q,T-n-1)$ degrees of freedom. For the likelihood--based statistics, if the values of the LR, W, and LM statistics are greater than $\chi_{1-\alpha}^2(q)$, then we reject hypothesis \eqref{eq36} at significance level $\alpha$, where $\chi_{1-\alpha}^2(q)$ is the $(1-\alpha)$ quantile of the chi--squared distribution with $q$ degrees of freedom. The same explanation can be made for the non--dividend paying company. 

\subsection{The Bayesian Estimation}

Now, we move to the Bayesian analysis of linear regression. In the Bayesian analysis, it assume that an analyst has a prior probability belief $f(\theta)$ about the unknown parameter vector $\theta=(\beta',\sigma^2)'$, where $f(\theta)$ is a prior density function of the vector $\theta$. Let us assume that prior density functions of the parameters $\beta$ and $\sigma^{-2}$, which is known as the precision of the parameter $\sigma^2$ are multivariate normal with mean $\beta_0$ and covariance matrix $\sigma^2B_0$ conditional on $\sigma^2$ and inverse--gamma distribution with parameters $\nu_0$ and $\lambda_0$, respectively, where for the dividend--paying company, dimensions of the mean $\beta_0$ and covaraince matrix $\sigma^2B_0$ are $((n+1)\times 1)$ and $((n+1)\times (n+1))$, respectively, while for the non--dividend paying company, the dimensions are $(n\times 1)$ and $(n\times n)$, respectively. Thus, the prior density functions are
\begin{equation}\label{}
f(\sigma^{-2})=\frac{(\lambda_0/2)^{\nu_0/2}\sigma^{-2(\nu_0/2-1)}\exp\{-\lambda_0\sigma^{-2}/2\}}{\Gamma(\nu_0/2)}
\end{equation}
and
\begin{equation}\label{ad05}
f(\beta|\sigma^{-2})=\frac{1}{(2\pi\sigma^2)^{(n+1)/2}|B_0|^{1/2}}\exp\bigg\{-\frac{1}{2\sigma^2}(\beta-\beta_0)'B_0^{-1}(\beta-\beta_0)\bigg\}.
\end{equation}
From the conditional density function in equation \eqref{ad05}, one can deduce that the analyst's best guess of the parameter $\beta$ is the vector $\beta_0$, and the confidence in this guess is summarized by the matrix $\sigma^2B_0$ and less confidence is represented by larger diagonal elements of $B_0$. 

After values of $b$ and $X$ is observed, the likelihood function $f(b|\theta,X)$ will update our beliefs about the parameter $\theta$. Which leads to a posterior density function $f(\theta|b,X)$. For each numerical value of the parameter $\theta$, the posterior density $f(\theta|b,X)$ describes our belief that $\theta$ is the true value, having observed values of $b$ and $X$. For the unknown parameter $\theta$, it can be shown (see \citeA{Hamilton94}) that its posterior density function is given by the following equation
\begin{equation}\label{eq49}
f(\theta|b,X)=f(\beta,\sigma^{-2}|b,X)=f(\beta|\sigma^{-2},b,X)f(\sigma^{-2}|b,X)
\end{equation}
where
\begin{equation}\label{eq50}
f(\beta|\sigma^{-2},b,X)=\frac{1}{(2\pi\sigma^2)^{(n+1)/2}|\bar{B}|^{1/2}}\exp\bigg\{-\frac{1}{2\sigma^2}(\beta-\bar{\beta})'\bar{B}^{-1}(\beta-\bar{\beta})\bigg\}
\end{equation}
with
\begin{equation}\label{eq51}
\bar{\beta}=\big(B_0^{-1}+X'X\big)^{-1}\big(B_0^{-1}\beta_0+X'b\big)~~~\text{and}~~~\bar{B}:=\big(B_0^{-1}+X'X\big)^{-1}
\end{equation}
and
\begin{equation}\label{eq52}
f(\sigma^{-2}|b,X)=\frac{(\bar{\lambda}/2)^{\bar{\nu}/2}\sigma^{-2(\bar{\nu}/2-1)}\exp\{-\bar{\lambda}\sigma^{-2}/2\}}{\Gamma(\bar{\nu}/2)}
\end{equation}
with 
\begin{eqnarray}\label{eq53}
\bar{\nu}&:=&\nu_0+T \nonumber\\
\bar{\lambda}&:=&\lambda_0+(b-X\hat{\beta})'(b-X\hat{\beta})+(\hat{\beta}-\beta_0)'B_0^{-1}\big(B_0^{-1}+X'X\big)^{-1}X'X(\hat{\beta}-\beta_0)
\end{eqnarray}
Note that if $B_0^{-1}\to 0$, which corresponds to uninformative diffuse prior, then the posterior mean \eqref{eq51} converges to the maximum likelihood estimator $\hat{\beta}= (X'X)^{-1}X'b$. Using the tower property of conditional expectation it follows the posterior mean equation \eqref{eq51} that the Bayesian estimator of the parameter vector $\beta$ is obtained by
\begin{equation}\label{}
\mathbb{E}(\beta|b,X)=\bar{\beta}=\big(B_0^{-1}+X'X\big)^{-1}\big(B_0^{-1}\beta_0+X'b\big).
\end{equation}
Due to the expectation formula of inverse--gamma distribution, and equations \eqref{eq52} and \eqref{eq53}, one can obtain that the Bayesian estimator of the precision $\sigma^{-2}$ is given by
\begin{equation}\label{}
\mathbb{E}(\sigma^{-2}|b,X)=\bar{\nu}/\lambda.
\end{equation}

To make statistical inference about the parameter vector $\theta=(\beta',\sigma^2)'$ conditional on the information $b$ and $X$, one may use the Gibbs sampling method, which generates a dependent sequence of our parameters. In the Bayesian statistics, the Gibbs sampling is often used when the joint distribution is not known explicitly or is difficult to sample from directly, but the conditional distribution of each variable is known and is easy to sample from. Constructing the Gibbs sampler to approximate the joint posterior distribution $f(\beta,\sigma^2|b,X)$ given in equation \eqref{eq49} is straightforward: New
values $\big(\beta_{(s)}, \sigma_{(s)}^2\big)$, $s=1,\dots,N$ can be generated by
\begin{itemize}
\item[1.] sample $\sigma_{(s)}^2\sim \mathcal{IG}(\bar{\nu},\bar{\lambda})$ 
\item[2.] sample $\beta_{(s)}\sim \mathcal{N}\big(\bar{\beta},\sigma_{(s)}^2\bar{B}\big),$
\end{itemize}
where $\mathcal{IG}$ is an abbreviation of the inverse--gamma distribution, and the parameters $\bar{\nu}$ and $\bar{\lambda}$ of the inverse--gamma distribution and mean $\bar{\beta}$ and covariance matrix $\bar{B}$ of the multivariate normal distribution are given in equations \eqref{eq51} and \eqref{eq53}, respectively. 

\subsection{Regime--Switching Estimation}

Now, we consider a private company valuation model with the regime--switching. In this case, the private company valuation models with $N$ regimes corresponding to equations \eqref{eq30} and \eqref{ad02} are given by the following equation 
\begin{equation}\label{ad08}
b_t=k_t^\circ(s_t)-\delta(s_t)\Delta_t+u_t=c_t'k(s_t)-\delta(s_t)\Delta_t+u_t, ~~~t=1,\dots,T
\end{equation}
and
\begin{equation}\label{}
b_t=k_t^\circ(s_t)+u_t=c_t'k(s_t)+u_t, ~~~t=1,\dots,T,
\end{equation}
respectively, where we assume that the price--to--book ratio and parameters of the required rate of return depend on regime--switching process $s_t$. For the private company valuation model with regime--switching, the vector of all parameters equals $\theta:=(k(1)',\dots,k(N)',m(1),\dots,m(N),\sigma^2,\rho',\text{vec}(P)')'$ for the dividend--paying company and $\theta:=(k(1)',\dots,k(N)',\sigma^2,\rho',$ $\text{vec}(P)')'$ for the non--dividend paying company. If we assume that the regime--switching process in regime $j$ at time $t$, then the conditional density functions of the dividend--to--book ratio at time $t$ of the companies are given by the following equations: for dividend paying company,
\begin{equation}\label{}
\eta_{tj}:=f(b_t|s_t=j,\Delta_t,c_t;\alpha)=\frac{1}{\sqrt{2\pi}\sigma}\exp\bigg\{-\frac{\big(b_t-c_t'k(j)+\delta(j)\Delta_t\big)^2}{2\sigma^2}\bigg\}
\end{equation}
and for non--dividend paying company,
\begin{equation}\label{}
\eta_{tj}:=f(b_t|s_t=j,c_t;\alpha)=\frac{1}{\sqrt{2\pi}\sigma}\exp\bigg\{-\frac{\big(b_t-c_t'k(j)\big)^2}{2\sigma^2}\bigg\}
\end{equation}
for $t=1,\dots,T$ and $j=1,\dots,N$, where $\alpha:=(k(1)',\dots,k(N)',m(1),\dots,m(N),\sigma^2)'$ is a parameter vector for the dividend--paying company, and $\alpha:=(k(1)',$ $\dots,k(N)',\sigma^2)'$ is a parameter vector for the non--dividend paying company, which differ from the vectors of all parameters $\theta$ by the initial probability vector $\rho$ and transition probability matrix $\mathsf{P}$.

Let us define the following vectors and matrix: $\bar{\Delta}_j:=\big(\bar{\Delta}_{1,j},\dots,\bar{\Delta}_{T,j}\big)'$ and $\bar{b}_j:=\big(\bar{b}_{1,j},\dots,\bar{b}_{T,j}\big)'$ are ($T\times 1)$ vectors, whose $t$--th elements are given by $\bar{\Delta}_{t,j}:=\Delta_t\sqrt{(z_{t|T})_j}$ and $\bar{b}_{t,j}:=b_t\sqrt{(z_{t|T})_j}$ and which are analogies, adjusted by the regime $j$ of the vectors $\Delta$ and $b$, respectively, and $\bar{C}_j':=\big[\bar{C}_{1,j}:\dots:\bar{C}_{T,j}\big]$ is an $(n \times T)$ matrix, whose $t$--th column is given by an $(n \times 1)$ vector $\bar{z}_{t,j}:=z_t\sqrt{\big(z_{t|T}\big)_j}$ and which is an analogy, adjusted by the regime $j$ of the matrix $C$. Then, one can obtain that for the dividend--paying company, maximum likelihood estimators of the model's parameters are given by
\begin{equation}\label{eq47}
\hat{\delta}(j)=-\frac{\bar{\Delta}_j'M_{\bar{C}_j}\bar{b}_j}{\bar{\Delta}_j'M_{\bar{C}_j}\bar{\Delta}_j},~~~\hat{k}(j)=(\bar{C}_j'\bar{C}_j)^{-1}\bar{C}_j'\big[\hat{\delta}(j)\bar{\Delta}_j+\bar{b}_j\big],~~~\hat{\sigma}^2=\frac{1}{T}\sum_{j=1}^Ne_j'e_j,
\end{equation}
where $e_j:=\bar{b}_j-\bar{c}_{t}'\hat{k}(j)+\hat{\delta}(j)\bar{\Delta}_j$ is an unrestricted residual vector corresponding to the regime $j$ and $M_{\bar{C}_j}:=I_T-\bar{C}_j(\bar{C}_j'\bar{C}_j)^{-1}\bar{C}_j'$ is a $(T\times T)$ symmetric idempotent matrix, and for non--dividend paying company, the parameter estimators are
\begin{equation}\label{ad06}
\hat{k}(j)=(\bar{C}_j'\bar{C}_j)^{-1}\bar{C}_j'\bar{b}_j,~~~\hat{\sigma}^2=\frac{1}{T}\sum_{j=1}^Ne_j'e_j,
\end{equation}
where $e_j:=\bar{b}_j-\bar{c}_{t}'\hat{k}(j)$ is a $(T\times 1)$ unrestricted residual vector corresponding to the regime $j$. It should be noted that for a dividend--paying company, the maximum likelihood estimators $\hat{m}(j)$ and $\hat{k}(j)$ for $j=1,\dots,N$ can be obtained by the least square method. To obtain the least square estimators, let us consider the following regression equation
\begin{equation}\label{ad07}
\bar{b}_t=\beta_1(j)\bar{c}_{1t,j}+\beta_2(j)\bar{c}_{2t,j}+\dots+\beta_n(j)\bar{c}_{nt,j}+\beta_{n+1}(j)\bar{\Delta}_{t,j}+u_t
\end{equation}
for $t=1,\dots,T,~j=1,\dots,N$, where for $i=1,\dots,n$, $\bar{c}_{it,j}:=c_{it}\sqrt{(z_{t|T})_j}$. We denote least square estimators of the linear regression equation by $b_1^*(j),\dots,b_{n+1}^*(j)$, which correspond to the parameters $\beta_1(j),\dots,\beta_{n+1}(j)$. Then, one can show that the maximum likelihood estimators $\hat{m}(j)$ and $\hat{k}(j)$ are equal to the following estimators, which are based on the least square method:
\begin{equation}\label{eq48}
\hat{m}(j)=-1/b_{n+1}^*(j)~~~\text{and}~~~\hat{k}_i(j)=b_i^*(j),~i=1,\dots,n.
\end{equation}
If we remove the term $\beta_{n+1}(j)\bar{\Delta}_{t,j}$ from the linear regression equation \eqref{ad07}, then one obtains ML and/or OLS parameter estimator for the non--dividend paying company. Again, the maximum likelihood estimator of the parameter vector $\theta$ is obtained by the zig--zag iteration method using the equations \eqref{eq41}--\eqref{eq43}, \eqref{eq45}, \eqref{eq47}/\eqref{eq48}, and \eqref{ad06}.

\subsection{The Kalman Filtering}

Now, we assume that the price--to--book ratio varies over time, that is, $m_t=P_t/B_t$, $t=1,\dots,T$. Under the assumption, equation \eqref{eq28} becomes
\begin{equation}\label{eq54}
m_tB_t=\big((1+k_t^\circ)m_{t-1}-\Delta_t\big)B_{t-1}.
\end{equation} 
Therefore, using the relation $B_t=(1+b_t)B_{t-1}$ in equation \eqref{eq54} a relation between the dividend--to--book ratio, book value growth rate, required rate of return, and price--to--book ratios is given by
\begin{equation}\label{eq55}
\Delta_t=-(1+b_t)m_t+(1+k_t^\circ)m_{t-1}.
\end{equation}
To estimate parameters of the required rate of return, we must add a random amount, say, $u_t$, into equation \eqref{eq55}. Then, equation \eqref{eq55} becomes
\begin{equation}\label{eq56}
\Delta_t=-(1+b_t)m_t+(1+k_t^\circ)m_{t-1}+u_t.
\end{equation}
It should be noted that for the above equation, the price--to--book ratios $m_t$ and $m_{t-1}$ are unobserved (state) variables. 
For a non--dividend--paying firm, the above equation becomes
\begin{equation}\label{}
\tilde{b}_t=\tilde{k}_t^\circ-\tilde{m}_t+\tilde{m}_{t-1}+u_t,
\end{equation}
where $\tilde{b}_t:=\ln(1+b_t)$ is a log book value growth rate, $\tilde{k}_t^\circ:=\ln(1+k_t^\circ)$ is a log required rate of return, and $\tilde{m}_t:=\ln(m_t)$ is an unobserved log price--to--book ratio, respectively, at time $t$ of the non--dividend paying company. Simplicity, we assume that the price--to--book ratio and log price--to--book ratio are governed by AR(1) process, that is, $m_t=\phi_0+\phi_1m_{t-1}+v_t$ and $\tilde{m}_t=\phi_0+\phi_1\tilde{m}_{t-1}+v_t$. Observe that if $\phi_1=1$, then the AR(1) process becomes the unit root process with drift. Consequently, our models are given by the following systems
\begin{equation}\label{eq57}
\begin{cases}
\Delta_t=-(1+b_t)m_t+(1+k_t^\circ)m_{t-1}+u_t\\
m_t=\phi_0+\phi_1m_{t-1}+v_t
\end{cases}~~~\text{for}~t=1,\dots,T
\end{equation}
for the dividend--paying company, and
\begin{equation}\label{ad09}
\begin{cases}
\tilde{b}_t=\tilde{k}_t^\circ-\tilde{m}_t+\tilde{m}_{t-1}+u_t\\
\tilde{m}_t=\phi_0+\phi_1\tilde{m}_{t-1}+v_t
\end{cases}~~~\text{for}~t=1,\dots,T
\end{equation}
for the non--dividend paying company. The systems \eqref{eq57} and \eqref{ad09} are more compactly written by
\begin{equation}\label{eq58}
\begin{cases}
y_t=\psi_t' z_t+\pi_t+u_t\\
z_t=Az_{t-1}+a+\eta_t
\end{cases}~~~\text{for}~t=1,\dots,T,
\end{equation}
where for the dividend--paying company, the endogenous variable $y_t$ equals the dividend--to--book ratio, $\Delta_t$, $z_t:=(m_t,m_{t-1})'$ is a ($2\times 1$) state vector of the price--to--book ratios at times $t$ and $t-1$, $\psi_t:=\big(-(1+b_t),1+k_t^\circ\big)'$ is a ($2\times 1$) vector, and $\pi_t:=0$ and for the non--dividend paying company, the endogenous variable $y_t$ equals the observed log book value growth rate, $\tilde{b}_t$, $z_t:=(\tilde{m}_t,\tilde{m}_{t-1})'$ is a ($2\times 1$) state vector of the log price--to--book ratios at times $t$ and $t-1$, $\psi_t:=(-1,1)'$ is a ($2\times 1$) vector, $\pi_t:=\tilde{k}_t^\circ$, and $a:=(\phi_0,0)'$ is a ($2\times 1$) vector, $\eta_t:=(v_t,0)'$ is a ($2\times 1$) random vector, and
$$A:=\begin{bmatrix}
\phi_1 & 0\\
1 & 0
\end{bmatrix}$$
is a ($2\times 2$) matrix. Note that system \eqref{eq58} can be easily extended to AR($p$) process for a transition equation (second line equation of the system), where $p$ is a natural number, and for the transition equation, the second components of the left and right--hand sides equal $m_{t-1}$ and $\tilde{m}_{t-1}$. 

The stochastic properties of systems \eqref{eq57}--\eqref{eq58} are governed by the random variables $u_1,\dots,u_T,$ $v_1,\dots,v_T$, $m_0$, and $\tilde{m}_0$. We assume that the error random variables $u_t$ and $v_t$ for $t=1,\dots,T$ and initial book--to--price ratio $m_0$ or log book--to--price ratio $\tilde{m}_0$ are mutually independent, and follow normal distributions, namely,
\begin{equation}\label{eq59}
m_0,\tilde{m}_0\sim \mathcal{N}(\mu_0,\sigma_0^2), ~~~u_t\sim \mathcal{N}(0,\sigma_u^2),~~~v_t\sim \mathcal{N}(0,\sigma_v^2), ~~~\text{for}~t=1,\dots,T.
\end{equation}

For the rest of the subsection, we review the Kalman filtering for our model, see also \citeA{Hamilton94} and \citeA{Lutkepohl05}. For $t=0,\dots,T$, let $x_t:=(y_t,z_t')'$ be a $(3\times 1)$ vector, composed of the endogenous variable $y_t$ and the state vector $z_t$, and $\mathcal{F}_t:=(B_0,\Delta_1,\dots,\Delta_t,b_1,\dots,b_t,c_1',\dots,c_t')'$ and $\mathcal{F}_t:=(B_0,\tilde{b}_1,\dots,\tilde{b}_t,c_1',\dots,c_t')'$ be a available information at time $t$ of dividend--paying and non--dividend paying companies, respectively. Then, system \eqref{eq58} can be written in the following form, which only depends on $z_{t-1}$
\begin{equation}\label{eq60}
x_t=\begin{bmatrix}
y_t \\ z_t
\end{bmatrix}=\begin{bmatrix}
\psi_t'a+\pi_t\\ a
\end{bmatrix}+\begin{bmatrix}
\psi_t'A \\ A
\end{bmatrix}z_{t-1}+\begin{bmatrix}
1 & \psi_t'\\
0 & I_2
\end{bmatrix}
\begin{bmatrix}
u_t \\ \eta_t
\end{bmatrix}~~~\text{for}~t=1,2,\dots.
\end{equation}
Because an error random vector $\zeta_t:=(u_t,\eta_t')'$ is independent of the information $\mathcal{F}_{t-1}$, conditional on $\mathcal{F}_{t-1}$, an expectation of a random vector $x_t:=(y_t,z_t)'$ is obtained by
\begin{equation}\label{eq61}
\begin{bmatrix}
y_{t|t-1}\\z_{t|t-1}
\end{bmatrix}:=\begin{bmatrix}
\mathbb{E}(y_t|\mathcal{F}_{t-1}) \\ \mathbb{E}(z_t|\mathcal{F}_{t-1})
\end{bmatrix}=\begin{bmatrix}
\psi_t'a+\pi_t\\ a
\end{bmatrix}+\begin{bmatrix}
\psi_t'A \\ A
\end{bmatrix}z_{t-1|t-1}
\end{equation}
for $t=1,\dots,T$, where $z_{0|0}:=(\mu_0,\mu_0)'$ is an initial value. If we use the tower property of conditional expectation and the fact that error random variables $u_t$ and $v_t$ are independent, and a error random vector $\zeta_t:=(u_t,\eta_t')'$ is independent of the information $\mathcal{F}_{t-1}$, then it is clear that
\begin{equation}\label{eq62}
\mathbb{E}\big((z_{t-1}-z_{t-1|t-1})\zeta_t'|\mathcal{F}_{t-1}\big)=0,~~~\mathbb{E}(u_t\eta_t'|\mathcal{F}_{t-1})=0,
\end{equation} 
for $t=1,\dots,T$. Consequently, it follows from equation \eqref{eq60} that conditional on $\mathcal{F}_{t-1}$, a covariance matrix of the random vector $x_t$ is given by
\begin{equation}\label{eq63}
\Sigma(x_t|t-1):=\text{Cov}(x_t|\mathcal{F}_{t-1})=\begin{bmatrix}
\Sigma(y_t|t-1) & \Sigma(z_t,y_t|t-1)'\\
\Sigma(z_t,y_t|t-1) & \Sigma(z_t|t-1)
\end{bmatrix}
\end{equation}
for $t=1,\dots,T$, where conditional on $\mathcal{F}_{t-1}$, a covariance matrix of the state vector $z_t$ is
\begin{equation}\label{eq65}
\Sigma(z_{t}|t-1)=A\Sigma(z_{t-1}|t-1)A'+\Sigma_\eta
\end{equation}
with $\Sigma_\eta:=\text{Cov}(\eta_t)=\text{diag}\{\sigma_v^2,0\}$,
conditional on $\mathcal{F}_{t-1}$, a variance of the endogenous variable $y_t$ is
\begin{equation}\label{eq64}
\Sigma(y_t|t-1):=\psi_t'\Sigma(z_{t}|t-1)\psi_t+\sigma_u^2
\end{equation}
with $\Sigma(z_{0}|0):=\text{diag}\{\sigma_0^2,\sigma_0^2\}$, and conditional on $\mathcal{F}_{t-1}$, a covariance matrix between the endogenous variable $y_t$ and the state vector $z_t$ is
\begin{equation}\label{eq66}
\Sigma(z_{t},y_t|t-1)=\Sigma(z_{t}|t-1)\psi_t
\end{equation}
As a result, due to equations \eqref{eq64}--\eqref{eq66}, for given $\mathcal{F}_{t-1}$, a conditional distribution of the process $x_t$ is given by
\begin{equation}\label{eq67}
x_t=\begin{bmatrix}
y_t \\ z_t
\end{bmatrix}
~\bigg|~\mathcal{F}_{t-1}\sim \mathcal{N}\left(\begin{bmatrix}
y_{t|t-1} \\ z_{t|t-1}
\end{bmatrix}, \begin{bmatrix}
\Sigma(y_t|t-1) & \Sigma(z_t,y_t|t-1)'\\
\Sigma(z_t,y_t|t-1) & \Sigma(z_t|t-1)
\end{bmatrix}
\right).
\end{equation}
It follows from the well--known formula of the conditional distribution of multivariate random vector and equation \eqref{eq67} that a conditional distribution of the state vector $z_t$ given the endogenous variable $y_t$ and the information $\mathcal{F}_{t-1}$ is given by
\begin{equation}\label{eq68}
z_t~|~y_t,\mathcal{F}_{t-1}\sim \mathcal{N}\left(z_{t|t-1}+\mathcal{K}_{t}\big(y_t-y_{t|t-1}\big),\Sigma(z_t|t-1)-\mathcal{K}_{t}\Sigma(y_t|t-1)\mathcal{K}_t'\right)
\end{equation}
for $t=1,\dots,T$, where $\mathcal{K}_t:=\Sigma(z_t,y_t|t-1)\Sigma^{-1}(y_t|t-1)$ is the Kalman filter gain. Therefore, since $\mathcal{F}_t=\{y_t,\mathcal{F}_{t-1}\}$, we have
\begin{equation}\label{eq69}
z_{t|t}:=\mathbb{E}(z_t|\mathcal{F}_t)=z_{t|t-1}+\mathcal{K}_{t}\big(y_t-y_{t|t-1}\big),~~~t=1,\dots,T
\end{equation}
and
\begin{equation}\label{eq70}
\Sigma(z_t|t):=\text{Cov}(z_t|\mathcal{F}_t)=\Sigma(z_t|t-1)-\mathcal{K}_{t}\Sigma(y_t|t-1)\mathcal{K}_t',~~~t=1,\dots,T.
\end{equation}

Because the error random vector $\eta_t:=(u_t,v_t)'$ for $t=T+1,T+2,\dots$ is independent of the full information $\mathcal{F}_T$ and the state vector at time $t-1$, $z_{t-1}$, it follows from equation \eqref{eq58} and the tower property of conditional expectation that Kalman filter's forecast step is given by the following equations 
\begin{equation}\label{eq71}
\begin{bmatrix}
y_{t|T}\\z_{t|T}
\end{bmatrix}=\begin{bmatrix}
\psi_t'z_{t|T}+\pi_t\\ Az_{t-1|T}+a
\end{bmatrix}~\text{and}~
\begin{bmatrix}
\Sigma(y_t|T)\\\Sigma(z_t|T)
\end{bmatrix}=\begin{bmatrix}
\psi_t'\Sigma(z_t|T)\psi_t+\sigma_u^2 \\ A\Sigma(z_{t-1}|T)A'+\Sigma_\eta
\end{bmatrix},~t=T+1,T+2,\dots.
\end{equation}

The Kalman filtering, which is considered the above provides an algorithm for filtering for the state vector $z_t$, which is the unobserved variable. To estimate parameters of our models \eqref{eq57} and \eqref{ad09}, in addition to the Kalman filter, we also need to make inference about the state vector $z_t$ for $t=1,\dots,T$ based on the full information $\mathcal{F}_T$, see below. Such an inference is called the smoothed estimate of the state vector $z_t$. The rest of the section is devoted to developing an algorithm, which is used to calculate the smoothed estimate $z_{t|T}:=\mathbb{E}(z_t|\mathcal{F}_T)$ for $t=0,\dots,T-1$.

Conditional on the information $\mathcal{F}_{t+1}$, a conditional distribution of a random vector $(z_t,z_{t+1})'$ is given by
\begin{equation}\label{eq73}
\begin{bmatrix}
z_{t+1} \\ z_t
\end{bmatrix}~\bigg|~\mathcal{F}_{t}\sim \mathcal{N}\left(
\begin{bmatrix}
z_{t+1|t} \\ z_{t|t}
\end{bmatrix},
\begin{bmatrix}
\Sigma(z_{t+1}|t) & \Sigma(z_t,z_{t+1}|t)'\\
\Sigma(z_t,z_{t+1}|t) & \Sigma(z_{t}|t)
\end{bmatrix}\right)
\end{equation}
for $t=0,\dots,T-1$, where $\Sigma(z_t,z_{t+1}|t):=\text{Cov}(z_t,z_{t+1}|\mathcal{F}_t)$ is a covariance between state vectors at times $t$ and $t+1$ given the information $\mathcal{F}_{t}$. It follows from equation \eqref{eq58} that the covariance is calculated by $\Sigma(z_t,z_{t+1}|t)=\Sigma(z_t|t)A'$. If we use the well--known formula of conditional distribution of multivariate random vector once again, then a conditional distribution of the random state vector at time $t$ given the state at time $t+1$ and the information $\mathcal{F}_{t}$ is given by
\begin{equation}\label{eq74}
z_t~|~z_{t+1},\mathcal{F}_{t}\sim \mathcal{N}\left(z_{t|t}+\mathcal{S}_{t}\big(z_{t+1}-z_{t+1|t}\big),\Sigma(z_t|t)-\mathcal{S}_t\Sigma(z_{t+1}|t)\mathcal{S}_t'\right)
\end{equation}
for $t=0,\dots,T-1$, where $\mathcal{S}_{t}:=\Sigma(z_t,z_{t+1}|t)\Sigma^{-1}(z_{t+1}|t)$ is the Kalman smoother gain. Because conditional on the state vector $z_{t+1}$, the state vector at time $t$, $z_t$, is independent of a endogenous variable vector $(y_{t+2},\dots,y_T)'$, for each $t=0,\dots,T-1$, it holds $\mathbb{E}(z_t|z_{t+1},\mathcal{F}_{T})=\mathbb{E}(z_t|z_{t+1},\mathcal{F}_t)=z_{t|t}+\mathcal{S}_{t}\big(z_{t+1}-z_{t+1|t}\big)$. Therefore, it follows from the tower property of the conditional expectation and conditional expectation in equation \eqref{eq74} that the smoothed inference of the state vector $z_t$ is obtained by
\begin{equation}\label{eq75}
z_{t|T}=\mathbb{E}\big(\mathbb{E}(z_t|z_{t+1},\mathcal{F}_T)\big|\mathcal{F}_T\big)=z_{t|t}+\mathcal{S}_{t}\big(z_{t+1|T}-z_{t+1|t}\big)
\end{equation}
for $t=0,\dots,T-1$. Using equation \eqref{eq75} a difference between the state vector $z_t$ and its Kalman smoother $z_{t|T}$ is represented by
\begin{equation}\label{eq76}
z_t-z_{t|T}=z_t-\big[z_{t|t}+\mathcal{S}_{t}(z_{t+1}-z_{t+1|t})\big]+\mathcal{S}_{t}(z_{t+1}-z_{t+1|T}).
\end{equation}
Observe that the square bracket term in the above equation is the conditional expectation of the state vector at time $t$, which is given in equation \eqref{eq74}. Thus, if we use conditional covariance matrix of the state vector $z_t$, which is given in equation \eqref{eq74} and use the tower property of conditional expectation once more, then we obtain that
\begin{equation}\label{eq77}
\Sigma(z_t|T)=\mathbb{E}\big((z_t-z_{t|T})(z_t-z_{t|T})'\big|\mathcal{F}_T\big)=\Sigma(z_t|t)-\mathcal{S}_{t}\big(\Sigma(z_{t+1}|t)-\Sigma(z_{t+1}|T)\big)\mathcal{S}_t'
\end{equation}
and
\begin{equation}\label{eq78}
\Sigma(z_{t},z_{t+1}|T)=\mathbb{E}\big((z_{t}-z_{t|T})(z_{t+1}-z_{t+1|T})'\big|\mathcal{F}_T\big)=\mathcal{S}_{t}\Sigma(z_{t+1}|T)
\end{equation}
for $t=0,\dots,T-1$. 

Let us consider the dividend--paying firm. In the EM algorithm, one considers a joint density function of a random vector, which is composed of observed variables and state (latent) variables. In our cases, the vectors of observed variables and state variables correspond to the vector of dividend--to--book ratios, $\Delta:=(\Delta_1,\dots,\Delta_T)'$, and a vector of price--to--book ratios, $m:=(m_0,\dots,m_T)'$, respectively. Interesting usages of the EM algorithm in econometrics can be found in \citeA{Hamilton90} and \citeA{Schneider92}. Let us denote the joint density function by $f_{\Delta,m}(\Delta,m)$. The EM algorithm consists of two steps. In the expectation (E) step of the EM algorithm, one has to determine the form of an expectation of log of the joint density given the full information $\mathcal{F}_T$. We denote the expectation by $\Lambda(\theta|\mathcal{F}_T)$, that is, $\Lambda(\theta|\mathcal{F}_T):=\mathbb{E}\big(\ln(f_{\Delta,m}(\Delta,m))|\mathcal{F}_T\big)$. For our model \eqref{eq57}, one can show that the expectation of log of the joint density of the vectors of the dividend--to--book ratios $\Delta$ and the price--to--book ratios $m$ is
\begin{eqnarray}\label{eq79}
&&\Lambda(\theta|\mathcal{F}_T)=\mathbb{E}\big(\ln(f_{\Delta,m}(\Delta,m))\big|\mathcal{F}_T\big)\nonumber\\
&&=-\frac{2T+1}{2}\ln(2\pi)-\frac{T}{2}\ln(\sigma_u^2)-\frac{1}{2\sigma_u^2}\sum_{t=1}^T\mathbb{E}\Big(\big(\Delta_t+(1+b_t)m_t-(1+k_t^\circ)m_{t-1}\big)^2\Big|\mathcal{F}_T\Big)\\
&&-\frac{T}{2}\ln(\sigma_v^2)-\frac{1}{2\sigma_v^2}\sum_{t=1}^T\mathbb{E}\Big(\big(m_t-\phi_0-\phi_1m_{t-1}\big)^2\Big|\mathcal{F}_T\Big)-\frac{1}{2}\ln(\sigma_0^2)-\frac{1}{2\sigma_0^2}\mathbb{E}\Big(\big(m_0-\mu_0\big)^2\Big|\mathcal{F}_T\Big),\nonumber
\end{eqnarray}
where $\theta:=\big(k',\phi_0,\phi_1,\mu_0,\sigma_u^2,\sigma_v^2,\sigma_0^2\big)'$ is an $((n+6)\times 1)$ vector, consisting of all parameters of model \eqref{eq57}. 

In the maximization (M) step of the EM algorithm, one needs to find a maximum likelihood estimator $\hat{\theta}$ that maximizes the expectation, which is determined in the E step. Taking partial derivatives from $\Lambda(\theta|\mathcal{F}_T)$ with respect to the parameters and setting these partial derivatives to zero gives the maximum likelihood estimators as
\begin{equation}\label{eq80}
\hat{k}:=\bigg(\sum_{t=1}^T\big[e_1'\Gamma(z_{t-1}|T)e_1\big]c_tc_t'\bigg)^{-1}\sum_{t=1}^Tz_t\Big(\Delta_te_1'z_{t-1|T}+(1+b_t)e_1'\Gamma(z_{t-1},z_t|T)e_1-e_1'\Gamma(z_{t-1}|T)e_1\Big),
\end{equation}
\begin{equation}\label{eq81}
\hat{\phi}_0:=\frac{\big[\sum_{t=1}^Te_1'z_{t|T}\big]\big[\sum_{t=1}^Te_1'\Gamma(z_{t-1}|T)e_1\big]-\big[\sum_{t=1}^Te_1'z_{t-1|T}\big]\big[\sum_{t=1}^Te_1'\Gamma(z_{t-1},z_t|T)e_1\big]}{T\big[\sum_{t=1}^Te_1'\Gamma(z_{t-1}|T)e_1\big]-\big[\sum_{t=1}^Te_1'z_{t-1|T}\big]^2},
\end{equation}
\begin{equation}\label{eq82}
\hat{\phi}_1:=\frac{T\big[\sum_{t=1}^Te_1'\Gamma(z_{t-1},z_t|T)e_1\big]-\big[\sum_{t=1}^Te_1'z_{t-1|T}\big]\big[\sum_{t=1}^Te_1'z_{t|T}\big]}{T\big[\sum_{t=1}^Te_1'\Gamma(z_{t-1}|T)e_1\big]-\big[\sum_{t=1}^Te_1'z_{t-1|T}\big]^2},
\end{equation}
\begin{equation}\label{eq83}
\hat{\mu}_0:=e_1'z_{0|0}, ~~~\hat{\sigma}_0^2:=e_1'\Sigma(z_0|0)e_1
\end{equation}
and
\begin{equation}\label{eq84}
\hat{\sigma}_u^2:=\frac{1}{T}\sum_{t=1}^T\mathbb{E}\big(u_t^2|\mathcal{F}_T\big), ~~~\hat{\sigma}_v^2:=\frac{1}{T}\sum_{t=1}^T\mathbb{E}\big(v_t^2|\mathcal{F}_T\big),
\end{equation}
where $e_1:=(1,0)'$ is a ($2\times 1$) unit vector, $\Gamma(z_{t-1}|T):=\mathbb{E}\big(z_{t-1}z_{t-1}'|\mathcal{F}_T\big)=\Sigma(z_{t-1}|T)+z_{t-1|T}z_{t-1|T}'$ is an  expectation of ($2\times 2$) dimensional random matrix $z_{t-1}z_{t-1}'$ given the full information $\mathcal{F}_T$, and $\Gamma(z_{t-1},z_t|T):=\mathbb{E}\big(z_{t-1}z_{t}'|\mathcal{F}_T\big)=\mathcal{S}_{t-1}\Sigma(z_{t}|T)+z_{t-1|T}z_{t|T}'$ is an expectation of ($2\times 2$) dimensional random matrix $z_{t-1}z_t$ given the full information $\mathcal{F}_T$.

To calculate the conditional expectations $\mathbb{E}\big(u_t^2|\mathcal{F}_T\big)$ and $\mathbb{E}\big(v_t^2|\mathcal{F}_T\big)$, let $u_{t|T}=y_t-\psi_t'z_{t|T}-\pi_t$ and $v_{t|T}=e_1'(z_{t|T}-Az_{t-1|T}-a)$ be smoothed residuals at time $t$ of the error random variables $u_t$ and $v_t$, respectively. It follows from equation \eqref{eq58} that 
\begin{eqnarray}\label{eq85}
u_t&=&u_{t|T}-\psi_t'(z_t-z_{t|T})\nonumber\\
v_t&=&v_{t|T}+e_1'(z_t-z_{t|T})-e_1'A(z_{t-1}-z_{t-1|T}).
\end{eqnarray}
Therefore, as $u_{t|T}$ and $v_{t|T}$ are known at time $T$ (measurable with respect to the full information $\mathcal{F}_T$), from equations \eqref{eq78}, one obtain that
\begin{eqnarray}\label{eq86}
\mathbb{E}\big(u_t^2|\mathcal{F}_T\big)&=&u_{t|T}^2+\psi_t'\Sigma(z_t|T)\psi_t\nonumber\\
\mathbb{E}\big(v_t^2|\mathcal{F}_T\big)&=&v_{t|T}^2+e_1'\Sigma(z_t|T)e_1+e_1'A\Sigma(z_{t-1}|T)A'e_1-2e_1'A\mathcal{S}_{t-1}\Sigma(z_{t}|T)e_1,
\end{eqnarray}
c.f. \citeA{Schneider92}. If we substitute equation \eqref{eq86} into \eqref{eq84}, then under suitable conditions the zig--zag iteration that corresponds to equations \eqref{eq61}, \eqref{eq65}, \eqref{eq64}, \eqref{eq69}, \eqref{eq70}, \eqref{eq75}, \eqref{eq77}, \eqref{eq80}--\eqref{eq84} converges to the maximum likelihood estimators of the dividend--paying company's parameters. 

Now we consider the non--dividend paying company. According to system \eqref{ad09}, conditional on the full information $\mathcal{F}_T$ corresponding to the non--dividend paying company, an expectation of log of the joint density of the log book value growth rates and the log price--to--book ratios is represented by the following equation
\begin{eqnarray}\label{}
&&\Lambda(\theta|\mathcal{F}_T)=-\frac{2T+1}{2}\ln(2\pi)-\frac{T}{2}\ln(\sigma_u^2)-\frac{1}{2\sigma_u^2}\sum_{t=1}^T\mathbb{E}\Big(\big(\tilde{b}_t-\tilde{k}_t^\circ+\tilde{m}_t-\tilde{m}_{t-1}\big)^2\Big|\mathcal{F}_T\Big)\\
&&-\frac{T}{2}\ln(\sigma_v^2)-\frac{1}{2\sigma_v^2}\sum_{t=1}^T\mathbb{E}\Big(\big(\tilde{m}_t-\phi_0-\phi_1\tilde{m}_{t-1}\big)^2\Big|\mathcal{F}_T\Big)-\frac{1}{2}\ln(\sigma_0^2)-\frac{1}{2\sigma_0^2}\mathbb{E}\Big(\big(\tilde{m}_0-\mu_0\big)^2\Big|\mathcal{F}_T\Big),\nonumber
\end{eqnarray}
where $\theta:=\big(k',\phi_0,\phi_1,\mu_0,\sigma_u^2,\sigma_v^2,\sigma_0^2\big)'$ is an $((n+6)\times 1)$ vector, which consists of all parameters of the model \eqref{eq57}. Taking a partial derivative from $\Lambda(\theta|\mathcal{F}_T)$ with respect to the parameter $k$ and setting to zero gives the maximum likelihood estimator as
\begin{equation}\label{}
\hat{k}:=\bigg(\sum_{t=1}^Tc_tc_t'\bigg)^{-1}\sum_{t=1}^Tc_t\Big(\tilde{b}_t+e_1'z_{t|T}-e_1'z_{t-1|T}\Big).
\end{equation}
The other parameters of the non--dividend paying company are given by equations \eqref{eq81}--\eqref{eq84}. Also, using same method as the dividend--paying company, one can obtain MLE of the company. It should be noted that the private company valuation model we consider in this section can be used not only by private companies but also by public companies.

\section{Numerical Results}

We start by applying the estimation method for parameter estimation of our model, see Section 3.4. For means of illustration, we have chosen three companies from different sectors (Healthcare, Financial Services, and Consumer), listed in the S\&P 500 index. In order to increase the number of price and dividend observation points, we take quarterly data instead of yearly data. Our data covers a period from Q1 1990 to Q3 2021. That leads to $T=126$ observations for Johnson \& Johnson, PepsiCo, and JPMorgan. All quarterly price and dividend data have been collected from Thomson Reuters Eikon.  

The dividends of the selected companies have different patterns. In particular, JPMorgan cut its dividend by a huge amount due to the 2008/2009 financial crises, and the other companies have continuously increasing dividend dynamics which are not affected by the 2008/2009 financial crises. For our model, we assume for all companies, that a default never occurs. 

We present estimations of the parameters for the selected companies in Table \ref{Tab1}. The 2--9th rows of Table 1 correspond to that the required rate of returns of the companies are modeled by the regime--switching process with three regimes and the 10--13th rows of the same Table correspond to that the required rate of returns of the companies take constant values (the regime--switching process takes one regime).

In order to obtain estimations of the parameters, which correspond to the 2--9th rows of Table 1 we assume that the regime--switching process $s_t$ follows a Markov chain with three regimes, namely, up regime (regime 1), normal regime (regime 2), and down regime (regime 3) and we used equations \eqref{eq43}--\eqref{eq46}. Since explanations are comparable for the other companies, we will give explanations only for PepsiCo. In the 2nd row of Table 1, we provide estimations of the parameters $k(1),k(2),k(3)$. For PepsiCo, in regimes 1, 2, and 3, estimations of the required rate of return are 11.39\%, 2.89\%, and --8.25\%, respectively. For example, in the normal regime, the required rate of return of PepsiCo could be 2.89\% on average. 

The 3--5th rows of Table 1 correspond to the transition probability matrix $P$. For the selected companies, their transition probability matrices $P$s are ergodic, where ergodic means that one of the eigenvalues of $P$ is unity and that all other eigenvalues of $P$ are inside the unit circle, see \citeA{Hamilton94}. From the 3rd row of Table 1 one can deduce that if the required rate of return of PepsiCo is in the up regime then in the next period, it will switch to the normal regime with a probability of 0.756 or the down regime with a probability of 0.244 because it can not be in the up regime due to zero probability. If the required rate of return of PepsiCo in the normal regime, corresponding to row 4 of the Table then in the next period, it will switch to the up regime with a probability of 0.077, the normal regime with a probability of 0.812, or the down regime with a probability of 0.111. Finally, if the required rate of return of PepsiCo is in the down regime then in the next period, it will switch to the up regime with a probability of 0.781 or the down regime with a probability of 0.111 because of the normal regime's zero probability, see 5th row of the same Table. 

We provide the average persistence times of the regimes in the 6th row of Table 1. The average persistence time of the regime $j$ is defined by $\tau_j:=1/(1-p_{jj})$ for $j=1,2,3.$ From Table 1, one can conclude that up, normal, and down regimes of PepsiCo's required rate of return will persist on average for 1.0, 5.3, and 1.3 quarters, respectively.

In the 7th row of Table 1, we give ergodic probabilities $\pi$ of the selected companies. Ergodic probability vector $\pi$ of an ergodic Markov chain is defined from an equation $P\pi=\pi$. The ergodic probability vector represents long--run probabilities, which do not depend on the initial probability vector $\rho=z_{1|0}$. After sufficiently long periods, the required rate of return of PepsiCo will be in the up regime with a probability of 0.169, the normal regime with a probability of 0.681, or the down regime with a probability of 0.150, which are irrelevant to initial regimes.

The 8th row of Table 1 is devoted to long--run expectations of the required rate of returns of the selected companies. The long--run expectation of the required rate of return is defined by $k_\infty:=\lim_{t\to \infty}\mathbb{E}(k(s_t)).$ For PepsiCo, it equals 2.66\%. So that after long periods, the average required rate of return of PepsiCo converges to 2.66\%.

\begin{table}[htbp]
  \centering
  \caption{Estimations of Parameters for the DDM}
    \begin{tabular}{|c|c|c|c|c|c|c|c|c|c|c|}
    \hline
    Row   & Prmtrs & \multicolumn{3}{c|}{Johnson \& Johnson} & \multicolumn{3}{c|}{PepsiCo} & \multicolumn{3}{c|}{JPMorgan} \\
    \hline
    2.     & $k(j)$  & 10.43\% & 4.72\% & --4.86\% & 11.39\% & 2.89\% & --8.25\% & 15.37\% & 5.21\% & --7.66\% \\
    \hline
    3.     & \multirow{3}{*}{$P$} & 0.000 & 1.000 & 0.000 & 0.000 & 0.756 & 0.244 & 0.397 & 0.000 & 0.603 \\
\cline{1-1}\cline{3-11}    4.     &       & 0.000 & 0.762 & 0.238 & 0.077 & 0.812 & 0.111 & 0.080 & 0.323 & 0.598 \\
\cline{1-1}\cline{3-11}    5.     &       & 0.492 & 0.000 & 0.508 & 0.781 & 0.000 & 0.219 & 0.290 & 0.710 & 0.000 \\
    \hline
    6.     & $\tau_j$   & 1.000 & 4.199 & 2.034 & 1.000 & 5.327 & 1.281 & 1.659 & 1.476 & 1.000 \\
    \hline
    7.     & $\pi$    & 0.138 & 0.581 & 0.281 & 0.169 & 0.681 & 0.150 & 0.232 & 0.393 & 0.375 \\
    \hline
    8.     & $k_\infty$     & \multicolumn{3}{c|}{2.82\%} & \multicolumn{3}{c|}{2.66\%} & \multicolumn{3}{c|}{2.74\%} \\
    \hline
    9.     & $\sigma_3$ & \multicolumn{3}{c|}{3.007} & \multicolumn{3}{c|}{3.238} & \multicolumn{3}{c|}{5.029} \\
    \hline
    10.    & $k$     & \multicolumn{3}{c|}{2.55\%} & \multicolumn{3}{c|}{2.39\%} & \multicolumn{3}{c|}{2.88\%} \\
    \hline
    11.    & $k_L$  & \multicolumn{3}{c|}{0.86\%} & \multicolumn{3}{c|}{0.28\%} & \multicolumn{3}{c|}{0.04\%} \\
    \hline
    12.    & $k_U$  & \multicolumn{3}{c|}{4.24\%} & \multicolumn{3}{c|}{4.51\%} & \multicolumn{3}{c|}{5.72\%} \\
    \hline
    13.    & $\sigma_1$ & \multicolumn{3}{c|}{7.262} & \multicolumn{3}{c|}{8.400} & \multicolumn{3}{c|}{9.219} \\
    \hline
    \end{tabular}%
  \label{Tab1}%
\end{table}%

In the 9th row of Table 1, we present parameter estimations of standard deviations of the error random variables $u_t$ for the selected companies. For PepsiCo, the parameter estimation equals 3.238. The 13th row of Table 1 corresponds to the parameter estimations of standard deviations, in which the required rate of returns of the companies are modeled by regime--switching process with one regime. For PepsiCo, the parameter estimation equals 8.4, where we used equation \eqref{eq08}. As we compare the 9th row and 13th row of the Table, we can see that the estimations that correspond to the regime--switching process with three regimes are significantly lower than the ones that correspond to the regime--switching process with one regime. 

Finally, the required rate of returns estimation at time Q2 2021 of the firms are presented in row ten of the Table, while the corresponding 95\% confidence intervals are included in rows 11 and 12 below. To calculate the required rate of returns estimation and confidence bands, we used equations \eqref{eq08} and \eqref{eq88}. Note that since the required rate of return estimation expresses the average quarterly return of the companies, we can convert them yearly based using a formula $(1+k)^4-1$. The Table further illustrates average returns (2.39\% for PepsiCo) and return variability, as the return is supposed to lie within the (0.28\%, 4.51\%) interval with a 95\% probability.

For the selected firms, it will be interesting to plot the probabilistic inferences with a return series. For each period $t=1,\dots,T$ and each firm, the probabilistic inferences are calculated by equation \eqref{eq41} and the return series are calculated by the formula $k_t:=(P_t+d_t)/P_{t-1}-1$. In Figure 1, we plotted the resulting series as a function of period $t$. In Figure 1, the left axis corresponds to the return series, while the right axis corresponds to the probabilistic inference series for each company. 

\begin{figure}[!h]\label{Fig1}
\centering
\caption{Returns VS Regime Probabilities of Selected Companies} 
\includegraphics[width=165mm]{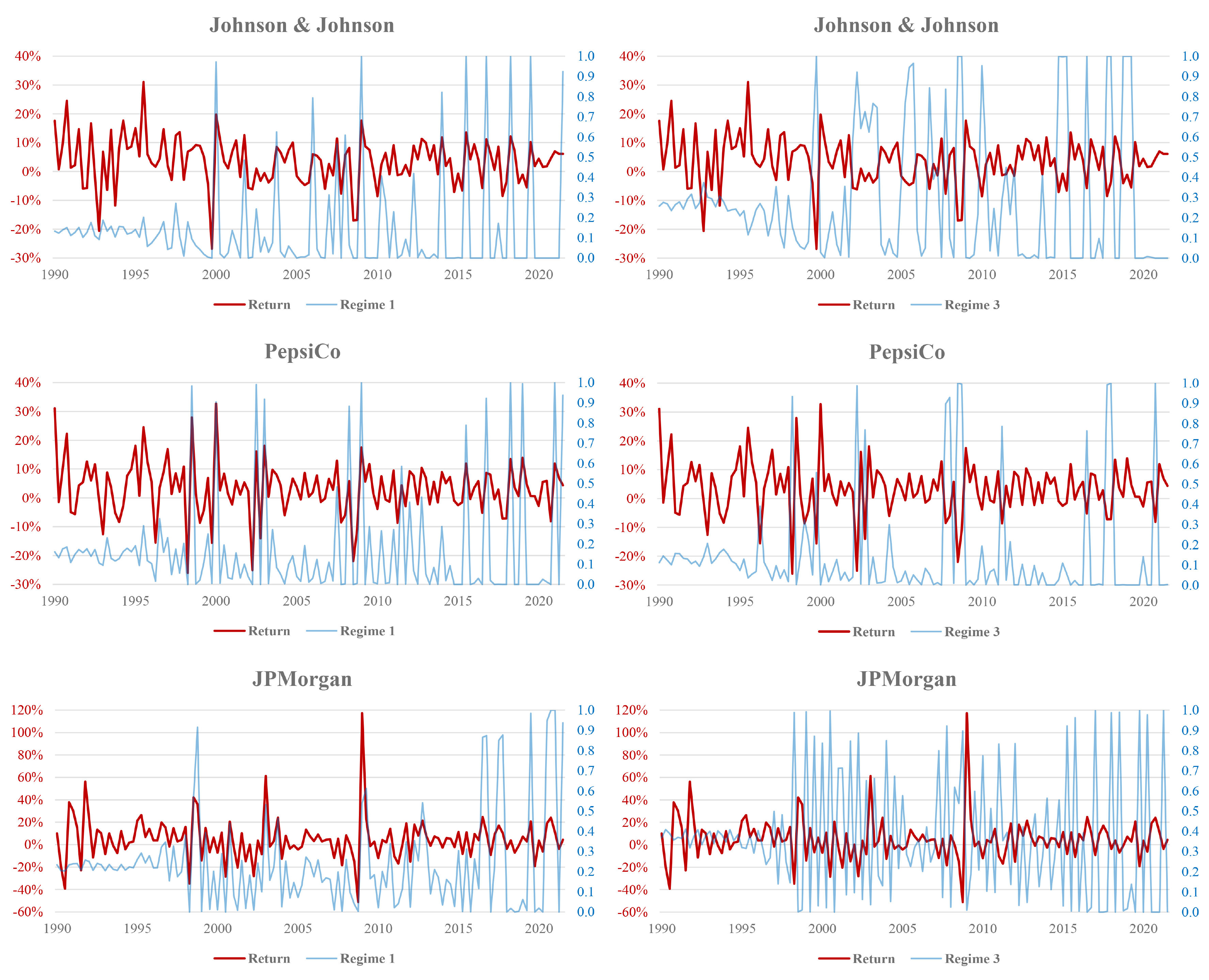} 
\end{figure}

From the Figure, and the 9th and 13th rows of Table 1, we can deduce that the regime--switching processes with three regimes are more suited to explain the required rate of return series as compared to the regime--switching processes with one regime.  

\section{Conclusion}

The required rate of return $k_t^\circ$ has some practical applications. For example, in addition to its usage in stock valuation,  it is an ingredient of the weighted average cost of capital (WACC), and WACC is used to value businesses and projects. Most popular practical method, which is used to estimate the required rate of return is the capital asset pricing model (CAPM). However, the CAPM is sensitive to its inputs. Therefore, in this paper, instead of the traditional CAPM and its descendant versions, we introduce new estimation methods, covering the ordinary ML methods, the Bayesian method, ML methods with regime--switching, and Kalman filtering to estimate the required rate of return. 

Our main purpose is the estimation of the required rate of return. However, the suggested methods can be used to estimate other parameters of the private company valuation model. In particular, we estimate of price--to--book ratio $m$ by the ordinary ML method, which is equivalent to the least square method and the Bayesian method, price--to--book ratios with regime--switching $m(1),\dots,m(N)$ by ML method with regime--switching, and state (unobserved and latent) variable of price--to--book ratio $m_t$ by Kalman filtering method. For the Kalman filtering method, we develop the EM algorithm. If we know the book values of the next periods one may use forecasting inferences of the state variable to value a company in the next periods. The parameter estimation methods of the private company valuation model can be used for not only by private companies but also by public companies.

It should be noted that the suggested methods can be used to estimate a required rate of return of debtholders. In particular, let $L_t$ be the market value of the liability of a company, $k_t^{\diamond}$ be the required rate of return of the debtholders, and $I_t$ be a payment at time $t$ of debts of the company, which includes interest payment. Then, equation \eqref{eq03} becomes
\begin{equation}\label{eq93}
L_t=(1+k_t^\diamond)L_{t-1}-I_t+u_t.
\end{equation}
Thus, one may estimate equation \eqref{eq93} by the suggested methods. Therefore, if we combine equations \eqref{eq03} and \eqref{eq93}, one can estimate the WACC of a company.

For the stochastic DDM, since we model the required rate of return $k_t^\circ$ by a linear function, depending on economic variables (covariates), i.e., $k_t^\circ=z_t'k$, our suggested methods can be used to explain movements of the stock price. For example, using dummy variables one may reveal the impact of the 2008/2009 financial crises on the stock prices of financial companies.

If the required rate of return satisfies $k_t^\circ<-1$, then according to equation \eqref{eq03}, the price at time $t$ takes a negative value, which is an undesirable result. Therefore, one may consider the following equation
\begin{equation}\label{eq92}
P_t=\exp\{\tilde{k}_t^\circ\}P_{t-1}-d_t+u_t,
\end{equation}
where $\tilde{k}_t^\circ=\ln(1+k_t^\circ)$ is the log required rate of return and can be modeled by the linear equation $\tilde{k}_t^\circ=k_1+k_2c_{2t}+\dots+k_nc_{nt}$. Thus, future research works should concentrate on the model given in equation \eqref{eq92}. Also, one may extend the private company valuation model with state variable by state--space model with regime--switching, see \citeA{Kim94}. Finally, the ideas in the paper can be extended to multiple stocks, which are dependent using panel regression.

\bibliographystyle{apacite}
\bibliography{References}

\end{document}